\documentclass[12pt,pre,pas,showpacs,reprint]{revtex4-1}
\usepackage{graphicx}
\usepackage{geometry}
\usepackage{mathrsfs}
\usepackage{subfigure}
\usepackage{amssymb}
\usepackage{indentfirst}
\usepackage{color}
\usepackage{cancel}
\usepackage{amsmath}
\usepackage{hyperref}
\usepackage[]{appendix}
\usepackage[]{color}
\usepackage{float}
\DeclareMathOperator\erf{erf}
\begin{document}
\title{\LARGE\textbf{Generation and Structural Characterization of Debye Random Media}}
\author{Zheng Ma}
\affiliation{Department of Physics, Princeton University, Princeton, New Jersey 08544, USA}

\author{Salvatore Torquato}
\email{torquato@electron.princeton.edu} 
\affiliation{Department of Chemistry, 
Department of Physics, Princeton Institute for the Science and Technology of Materials,\\
and Program in Applied and Computational Mathematics, Princeton University,
Princeton, New Jersey 08544, USA}

\begin{abstract}
In their seminal paper on scattering by an inhomogeneous solid, Debye and coworkers proposed a simple exponentially decaying function for the two-point correlation function of an idealized class of two-phase random media. Such {\it Debye random media}, which have been shown to be realizable, are singularly distinct from all other models of two-phase media in that they are entirely defined by their one- and two-point correlation functions. To our knowledge, there has been no determination of other microstructural descriptors of Debye random media. In this paper, we generate Debye random media in two dimensions using an accelerated Yeong-Torquato construction algorithm. We then ascertain microstructural descriptors of the constructed media, including their surface correlation functions, pore-size distributions, lineal-path function, and chord-length probability density function. Accurate semi-analytic and empirical formulas for these descriptors are devised. We compare our results for Debye random media to those of other popular models (overlapping disks and equilibrium hard disks), and find that the former model possesses a wider spectrum of hole sizes, including a substantial fraction of large holes. Our algorithm can be
applied to generate other models defined by their two-point correlation functions, and their other microstructural descriptors can be determined and analyzed by the procedures laid out here.          
\end{abstract} 
\date{}
\maketitle

\section{Introduction}
Disordered two-phase heterogeneous media are ubiquitous; examples include
composites, porous media, colloids, polymer blends, and biological media \cite{torquato2013random,Mi02,sahimi2003heterogeneous,Pa16,hristopulos2020random, gibson1999cellular, wadsworth2016universal}.
The microstructure of a two-phase medium can be completely statistically characterized
by an infinite set of $n$-point correlation functions
(defined in Sec. II) \cite{torquato2013random}. Interestingly, there exist unique models
of two-phase media for which one can explicitly represent and compute, in principal,
any $n$-point correlation function \cite{To86i,St95,torquato2013random}. What has come to be known as {\it Debye random media}
\cite{yeong1998reconstructing} are singularly distinct from all other two-phase models
in that they are entirely defined by their one- and two-point correlation functions (see Sec. III). Specifically, Debye et al. \cite{debye1957scattering} proposed an autocovariance function that is a simple exponentially
decaying function (see Eq.(\ref{S2Debye})) to model media with phases of ``fully random shape, size, and distribution."
Importantly, such autocovariance functions approximate well  those for realistic two-phase
media \cite{debye1957scattering}, including Fontainebleau sandstones \cite{coker1996morphology}.  

To our knowledge, there has been no determination 
of other microstructural descriptors of Debye random media. In this paper, we ascertain other descriptors of a certain class of Debye random media, including their surface correlation functions, pore-size distributions, lineal-path function, and chord-length probability density function. We accomplish
this program by generating 
Debye random media in two dimensions at five different volume fractions using the Yeong-Torquato construction procedure \cite{yeong1998reconstructing}
and then sampling for the aforementioned microstructural
descriptors. We also compare these different
descriptors for Debye random media to corresponding quantities for other models of two-phase media that have been commonly studied, including
dispersions of overlapping particles and equilibrium hard particles.

In Sec. II, we provide definitions of all of the microstructural descriptors
considered in this paper. In Sec. III, we define and discuss Debye random media. In Sec. IV, we provide
details of the accelerated Yeong-Torquato construction algorithm used to construct Debye random media. In Sec. V, we present results of the microstructural descriptors computed from our constructions. In Sec. VI, we compare these microstructural descriptors with those of popular models of particle dispersions. In Sec. VII, we make concluding remarks
and discuss possible
future research directions. 

\section{Definitions of Microstructural Descriptors}
\subsection{$n$-Point Correlation Function}
\indent Here we define several microstructural descriptors that are widely used to characterize random media. In general, a two-phase random medium is a domain of space $\mathcal V \subseteq \mathbb{R}^d$ that is partitioned into two disjoint regions that make up $\mathcal V$: a phase 1 region $\mathcal V_1$ of volume fraction $\phi_1$ and a phase 2 region $\mathcal V_2$ of volume fraction $\phi_2$ \cite{torquato2013random}. 

The phase indicator function $\mathcal I^{(i)}(\mathbf x)$ for a given realization is defined as
\begin{equation} \label{indicator}
\mathcal I^{(i)}(\mathbf x)=\left \{ \begin{aligned} & 1, & \mathbf x \in \mathcal V_i,\\ & 0, & \mathbf x \notin \mathcal V_i. \end{aligned}\right.
\end{equation}
Most generally, the $n$-point correlation function $S_n^{(i)}$ for phase $i$ \cite{torquato2013random} is defined as
\begin{equation} \label{sn}
S_n^{(i)}(\mathbf x_1,\mathbf x_2,...,\mathbf x_n)=\left\langle \prod_{i=1}^n\mathcal I^{(i)}(\mathbf x_i)\right \rangle.
\end{equation} 
The function has a probabilistic interpretation: It gives the probability of finding the ends of the vectors $\mathbf x_1$,...,$\mathbf x_n$ all in phase $i$. In this formalism, the volume fraction $\phi_{i}$ for phase $i$ is the one-point correlation function
\begin{equation}
S_1^{(i)}({\mathbf x}_1)=\left\langle \mathcal I^{(i)}(\mathbf x)\right\rangle,
\end{equation}
which is equal to the phase volume fraction $\phi_i$
(a constant) for statistically homogeneous media.

The commonly used two-point correlation function is written as 
\begin{equation} \label{s2}
S_2^{(i)}(\mathbf x_1,\mathbf x_2)=\left\langle \mathcal I^{(i)}(\mathbf x_1)\mathcal I^{(i)}(\mathbf x_2)\right\rangle.
\end{equation} 
For statistically homogeneous media, this quantity only depends on the relative displacement vector $\mathbf r\equiv \mathbf x_2-\mathbf x_1$. The two-point correlation function simplifies as $S_2(\mathbf x_1,\mathbf x_2)=S_2(\mathbf r)$. If the system is also statistically isotropic, then $S_2(r)$ depends only on the radial distance $r =|\bf{r}|$. The two-point correlation function $S_2^{(i)}(r)$ is related to the autocovariance function $\chi_{_V}(r)$ simply by subtracting its large-$r$ value, i.e., 
\begin{equation}
\chi_{_V}(r)\equiv S_2^{(1)}(r)-\phi_1^2=S_2^{(2)}(r)-\phi_2^2.
\end{equation}
Specifically, we have 
\begin{equation}
\lim_{r\rightarrow 0}\chi_{_V}(r)=\phi_1\phi_2, \lim_{r\rightarrow \infty}\chi_{_V}(r)=0,
\label{eq:chivlimit}
\end{equation}
the later holds when there is no long-range order. Note that $\chi_{_V}(r)$ is invariant to the choice of the phase. The Fourier transform of the autocovariance function is called spectral density $\tilde \chi_{_V}(k)$, which is another important quantity, and can be obtained from
scattering experiments \cite{debye1957scattering, teubner1990scattering}. \\
\indent An interesting property of the two-point correlation function shown by Debye and coworkers \cite{debye1957scattering} is that its derivative at the origin is proportional to the specific surface $s$ for three-dimensional isotropic media, which can be used to retrieve such information from scattering experiments. This property is further generalized to anisotropic media \cite{berryman1987relationship} as well as media in $d$ dimensions \cite{torquato2013random}, which writes as   
\begin{equation} \label{s2dr}
\frac{dS_2^{(i)}}{dr}\Bigr|_{r=0}=-\frac{\omega_{d-1}}{\omega_d d}s,
\end{equation}
where 
\begin{equation} \label{omega}
\omega_d=\frac{\pi^{d/2}}{\Gamma(1+d/2)}
\end{equation}
is the $d$-dimensional volume of a sphere of unit radius. In two and three dimensions, the derivative in Eq. (\ref{s2dr}) are simply $-s/\pi$ and $-s/4$, which we will apply in the following section.

\subsection{Surface Correlation Functions}
Equally important, but less well-known descriptors are the two-point surface correlation functions, which arise in rigorous
bounds on transport properties of porous media \cite{doi1976new, torquato2013random}. We first define the interface indicator function \cite{torquato2013random}
\begin{equation} \label{indicator1}
\mathcal M(\mathbf x)=|\nabla \mathcal I^{(1)}(\mathbf x)|=|\nabla \mathcal I^{(2)}(\mathbf x)|.
\end{equation} 
The specific surface is the expected area of the interface per unit volume, and for homogeneous media is simply the ensemble average of the interface indicator function, i.e.,
\begin{equation} \label{sss}
s=\left\langle \mathcal M(\mathbf x) \right\rangle.
\end{equation} 
\indent The surface-void correlation function $F_{sv}(\mathbf r)$ measures the correlation between one point on the interface and the other in the void phase. For homogeneous media, it is defined as 
\begin{equation} \label{deffsv}
F_{sv}(\mathbf r)=\left\langle \mathcal M(\mathbf x)\mathcal I^{(void)}(\mathbf x+\mathbf r) \right\rangle.
\end{equation}
Henceforth, we will denote phase 1 as the void phase while phase 2 as the solid phase. Similarly to the two-point correlation function, $F_{sv}(r)$ also has interesting small-$r$ behavior. Specifically, we previously showed that \cite{ma2018precise}
\begin{equation} \label{fsvdr}
F_{sv}(r)=s(\frac{1}{2}+\frac{r}{2B(\frac{d-1}{2},\frac{1}{2})} \bar H),
\end{equation}
where $B(\frac{d-1}{2},\frac{1}{2})$ is the beta function and $\bar H$ is the \textit{integrated mean curvature} $H$ averaged on the interface. Specifically, this implies that in two dimensions, the derivative of $F_{sv}(r)$ is related to the Euler characteristic $\chi$ by 
\begin{equation} \label{fsvdr2}
\frac{dF_{sv}(r)}{dr}\Bigr|_{r=0}=\frac{\chi}{V},
\end{equation}  
where the right hand side can be understood as an intensive property, or specific Euler characteristic. Apparently, as $r\rightarrow \infty$, we have 
\begin{equation}
\lim_{r\rightarrow \infty}F_{sv}(r)=s\phi_1,
\label{eq:fsvlimit}
\end{equation}
when there is no long-range order.

The surface-surface correlation function $F_{ss}(\mathbf r)$ measures the correlation between two points on the interface. For homogeneous media, it is defined as
\begin{equation} \label{deffss}
F_{ss}(\mathbf r)=\left\langle \mathcal M(\mathbf x)\mathcal M(\mathbf x+\mathbf r) \right\rangle.
\end{equation}
It can be shown that at small $r$, $F_{ss}(r)$ diverges as $(d-1)\omega_{d-1}s/d\omega_dr$ \cite{ma2018precise}. While as $r\rightarrow \infty$, we have 
\begin{equation}
\lim_{r\rightarrow \infty}F_{ss}(r)=s^2,
\label{eq:fsslimit}
\end{equation}
when there is no long-range order.

\subsection{Pore-Size Functions}
One important way to characterize the pore (void) space is by the pore-size probability density function $P(\delta)$, which is defined by \cite{torquato2013random}
\begin{equation}
P(\delta)=-\partial F(\delta)/\partial \delta, 
\label{eq:pore-dist}
\end{equation}
where $F(\delta)$ is the complementary cumulative distribution function that measures the probability that a randomly placed sphere of radius $\delta$ centered in the pore space $\mathcal V_1$ lies entirely in $\mathcal V_1$. Clearly, $F(0)=1$ and $F(\infty)=0$. Consequently, we have $P(0)=s/\phi_1$ and $P(\infty)=0$. The $n$th moment of the pore-size probability density is defined by \cite{torquato2013random}
\begin{eqnarray}
\langle \delta^n \rangle \equiv &\int_{0}^{\infty}\delta^n P(\delta)d\delta \nonumber \\
=&n\int_{0}^{\infty} \delta^{n-1}F(\delta)d\delta.
\label{eq:moments}
\end{eqnarray}
The moments of the pore-size probability density provide a measure of the characteristic length scale of the pore space, which has been shown to be useful in predicting transport properties of random media \cite{prager1961viscous, avellaneda1991rigorous}. In this paper, we are particularly interested in the first moment, i.e., the mean pore size $\langle \delta \rangle$, which we compute in Sec. V. 

\subsection{Lineal-Path Function}
Another interesting statistical descriptor that we consider in this paper is the lineal-path function $L^{(i)}(z)$ \cite{lu1992lineal}. The lineal-path function $L^{(i)}(z)$ is the probability that a line segment of length $z$ is entirely in phase $i$. This function provides degenerate connectedness information along a lineal path in phase $i$. Clearly, it is a monotonically decreasing function with $L^{(i)}(0)=\phi_i$ and $L^{(i)}(\infty)=0$. In Sec. V, we calculate $L(z)  \equiv L^{(1)}(z)$ to characterize the pore space of Debye random media. 

\subsection{Chord-Length Probability Density Function}
The chord-length probability density function $p^{(i)}(z)$ is another descriptor that is closely related to the lineal-path function $L^{(i)}(z)$ \cite{matheron1975random,torquato1993chord}. Here chords refer to all of the line segments between intersections of an infinitely long line with the two-phase interface. For statistically isotropic media, $p^{(i)}(z)dz$ is the probability of finding a chord of length between $z$ and $z+dz$ in phase $i$. The chord-length density function is of importance in the study of a variety of transport properties of porous media \cite{ho1979asymptotic,tokunaga1985porous,thompson1987microgeometry}. 

Interestingly, it has been shown that $p^{(i)}(z)$ is directly related to the second derivative
of the lineal-path function $L^{(i)}(z)$ \cite{torquato1993chord}, specifically, 
\begin{equation}
p^{(i)}(z)=\frac{\ell^{(i)}_C}{\phi_i}\frac{d^2L^{(i)}(z)}{dz^2},
\label{eq:pzlz}
\end{equation}
where $\ell^{(i)}_C$ is the mean chord length for phase $i$, i.e., $\ell^{(i)}_C=\int_{0}^{\infty}zp^{(i)}(z)dz$. For statistically isotropic systems, the mean chord length is related to the slope of the two-point correlation function at the origin via the expression 
\begin{equation}
\ell^{(i)}_C=\frac{\phi_i}{-\frac{dS_2^{(i)}}{dr}\Bigr|_{r=0}}=\frac{\omega_d\phi_id}{\omega_{d-1}}\frac{1}{s}.
\label{eq:chords2}
\end{equation} 
In two and three dimensions, $\ell^{(i)}_C$ is simply given by $\pi\phi_i/s$ and $4\phi_i/s$, respectively \cite{underwood1970quantitative}. In Sec. V, we calculate $p(z)  \equiv p^{(1)}(z)$ to characterize the pore space of Debye random media.  

\section{Debye Random Media}
\begin{figure*}[]
\centering
\subfigure[]{
\includegraphics[width=5cm, height=5cm]{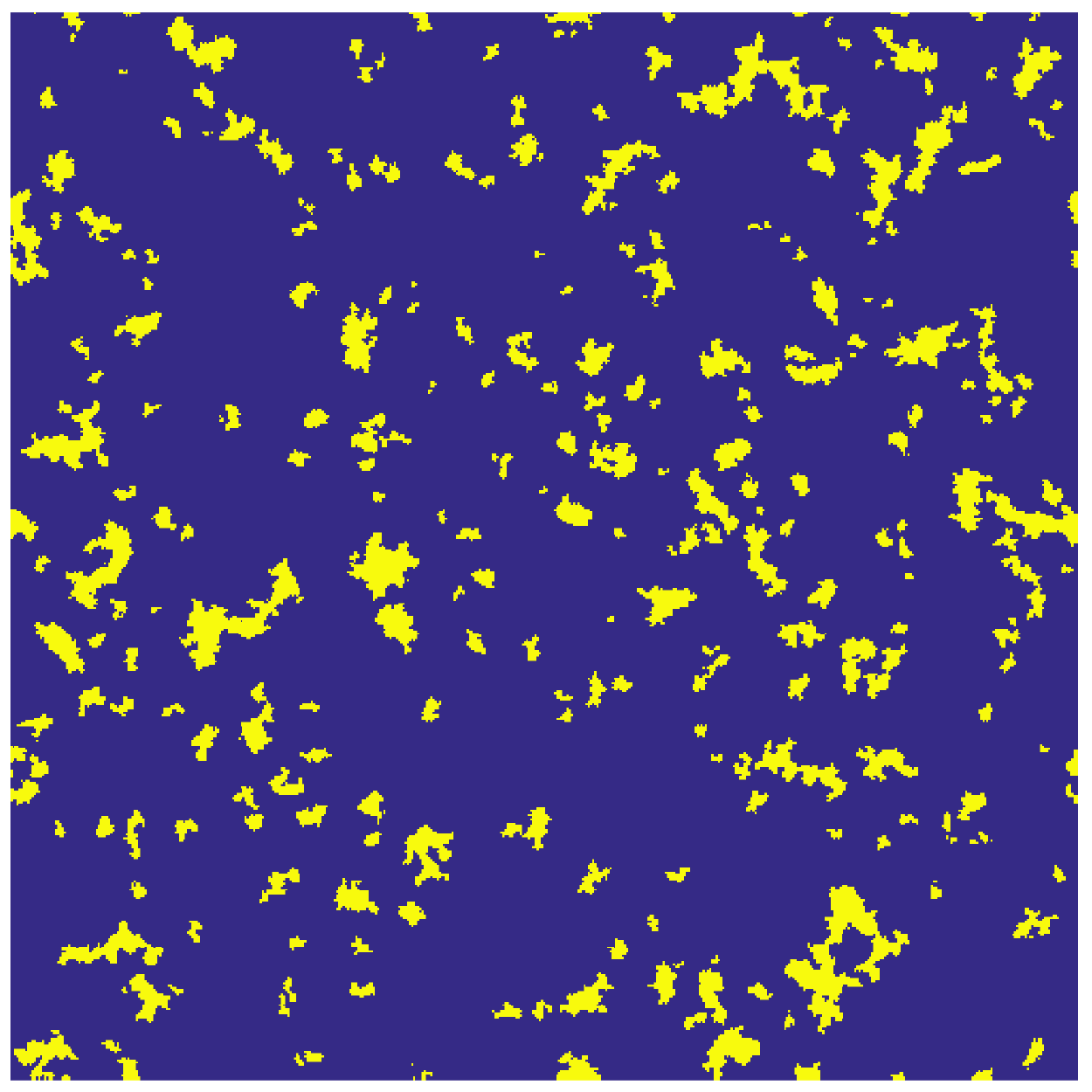}
}
\subfigure[]{
\includegraphics[width=5cm, height=5cm]{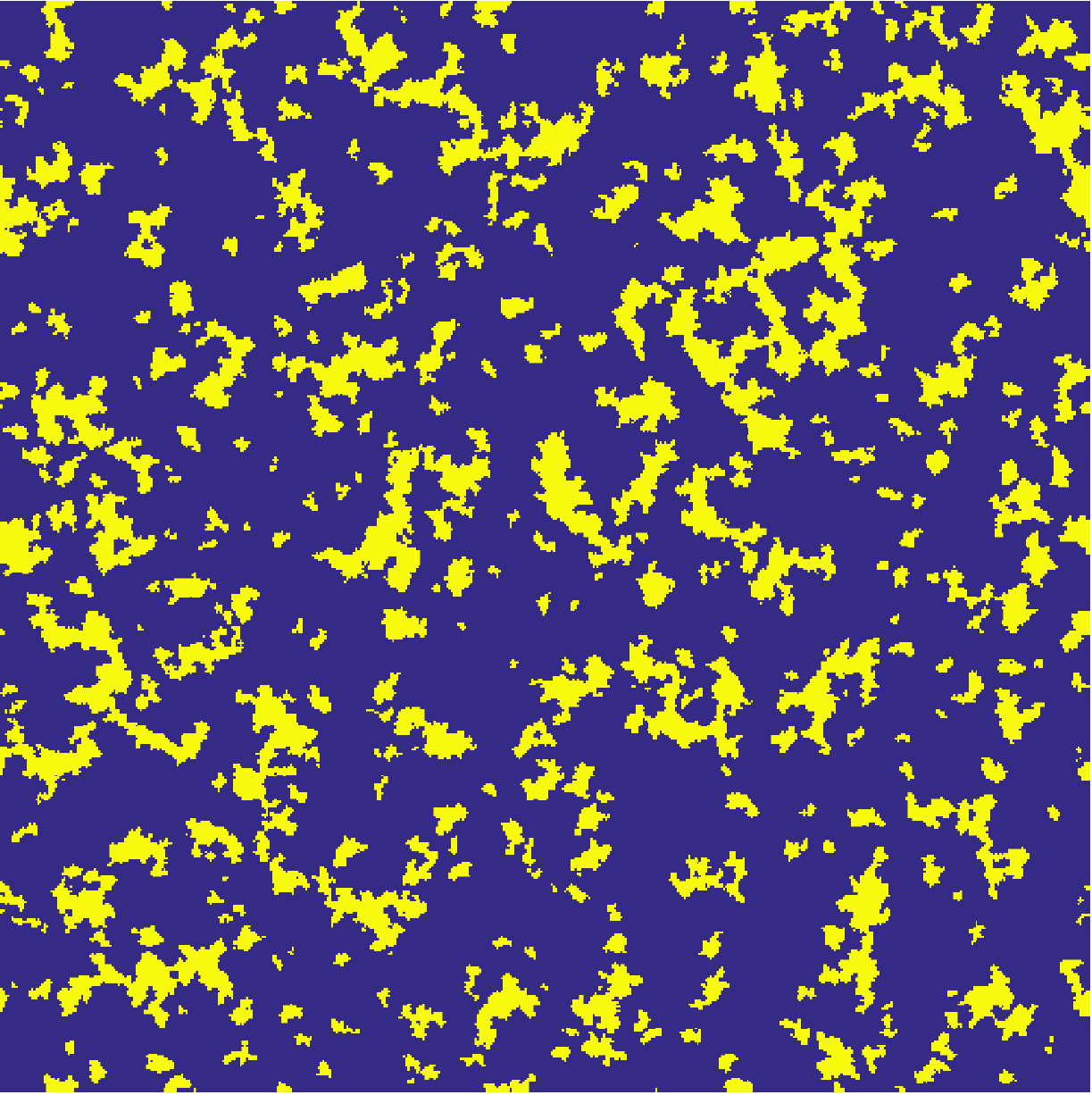}
}
\subfigure[]{
\includegraphics[width=5cm, height=5cm]{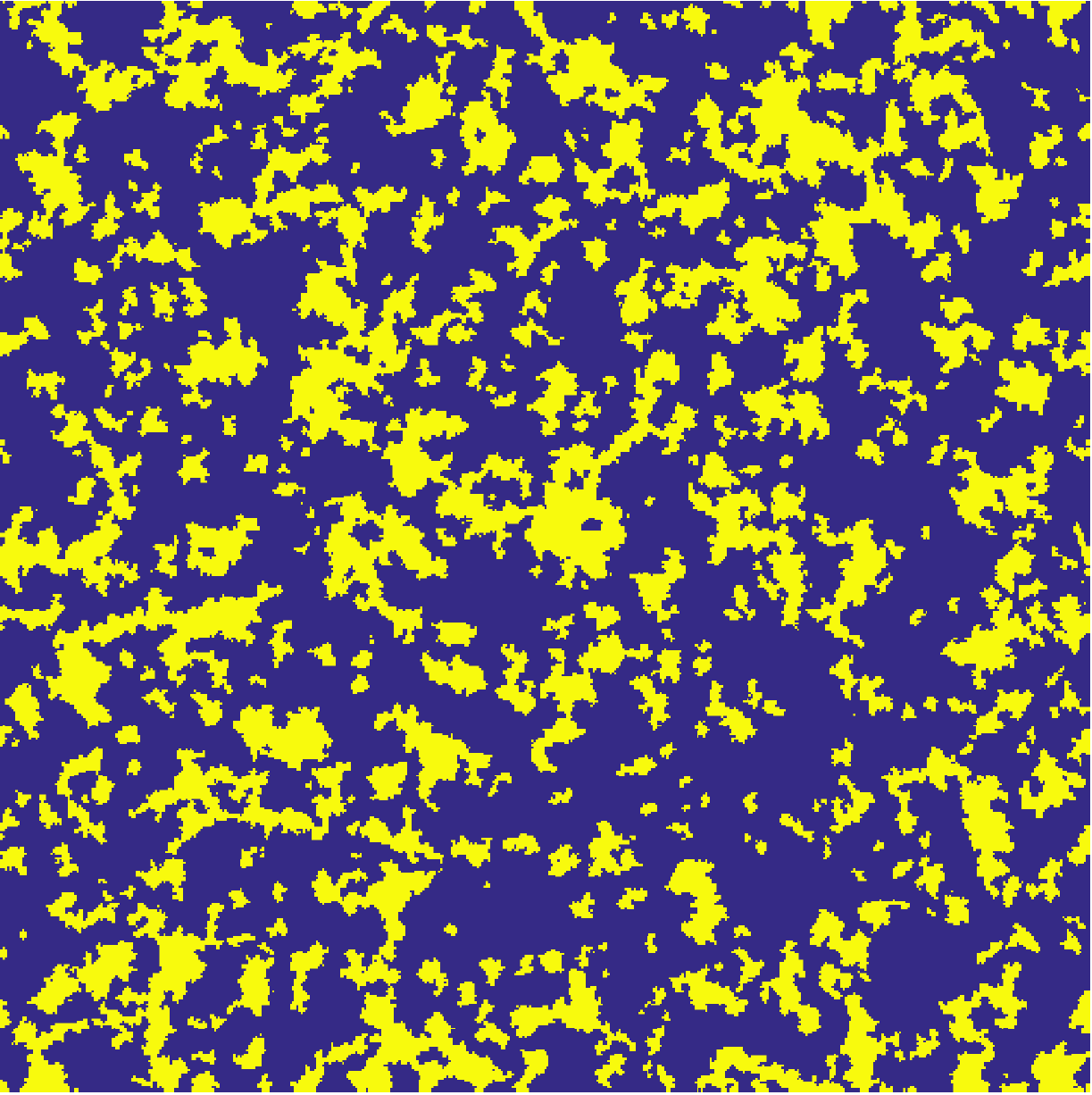}
}
\subfigure[]{
\includegraphics[width=5cm, height=5cm]{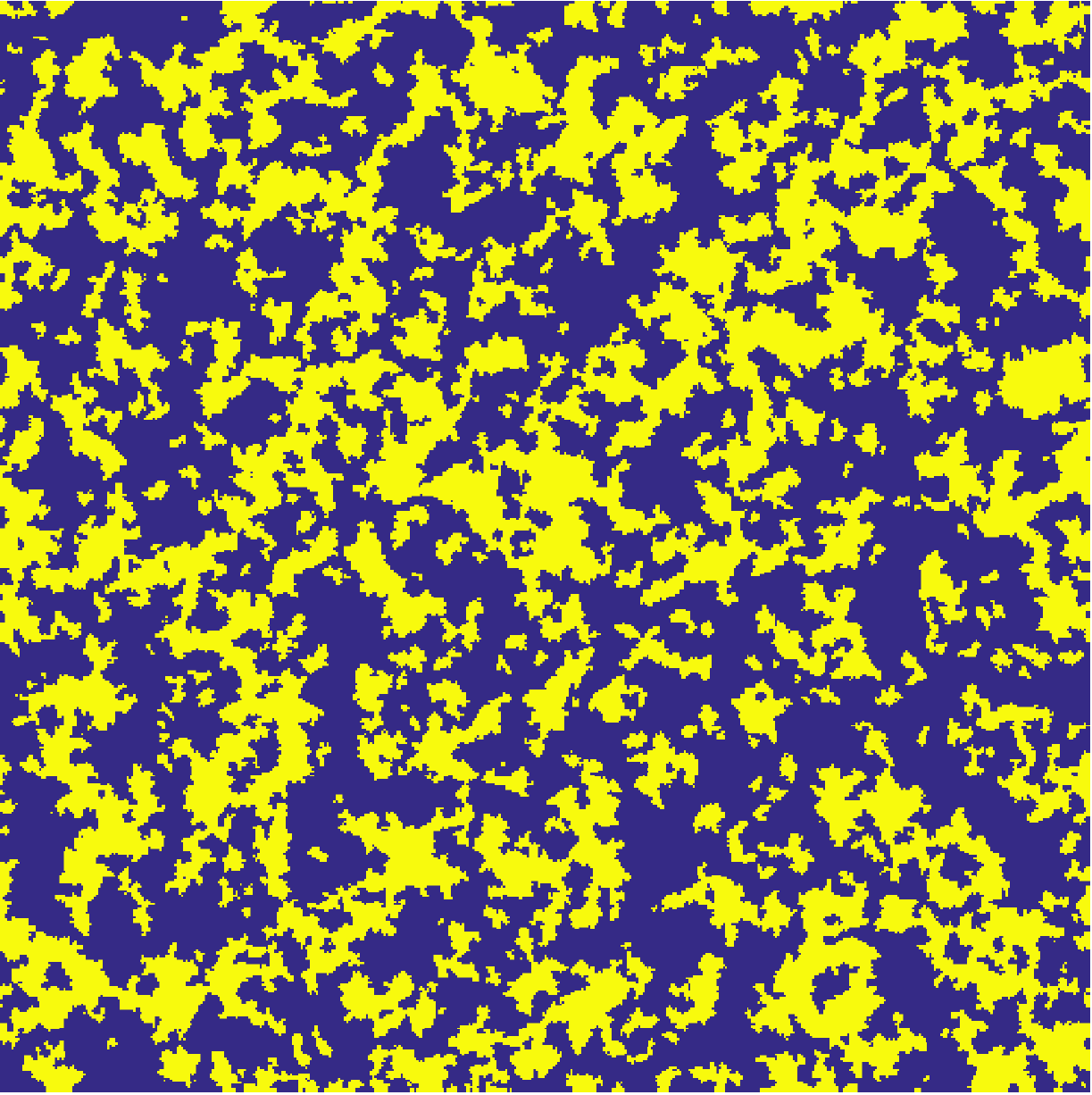}
}
\subfigure[]{
\includegraphics[width=5cm, height=5cm]{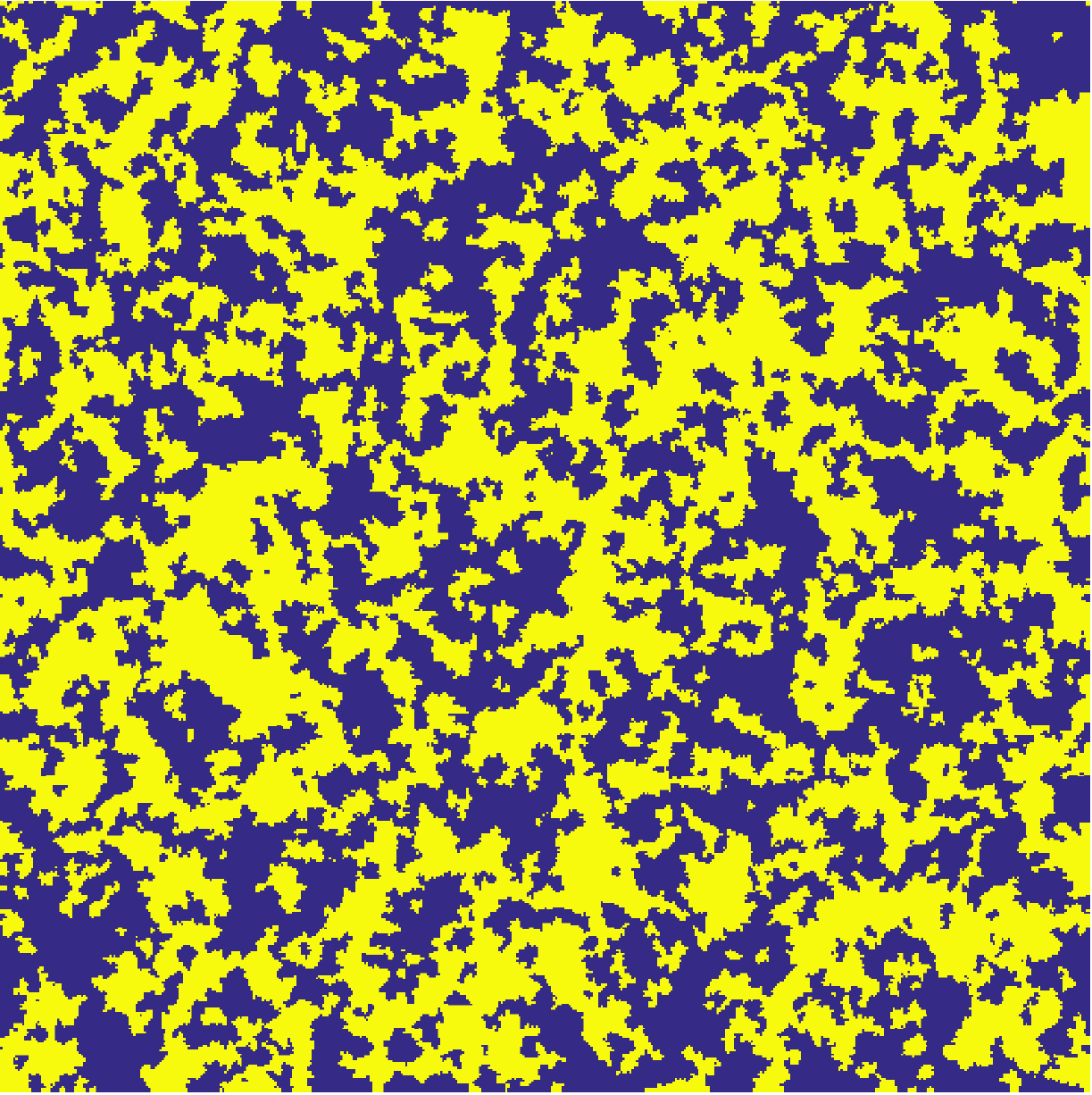}
}
\caption{Representative digitized images of Debye random media obtained
from the construction algorithm at different volume fractions. Each sample consists of $N^2=501^2$ pixels
and the length scale $a=5$ pixels. The volume fractions of the solid (yellow) phase are (a) $\phi_2=0.1$. (b) $\phi_2=0.2$. (c) $\phi_2=0.3$. (d) $\phi_2=0.4$. (e) $\phi_2=0.5$.}
\label{fig: reconstruct}
\end{figure*}
Debye random media in $d$-dimensional space 
$\mathbb{R}^d$ are a class of statistically homogeneous and isotropic two-phase media that is entirely
defined by its radial two-point correlation function or, equivalently,
its autocovariance function \cite{debye1957scattering, yeong1998reconstructing}:

\begin{equation} \label{S2Debye}
\chi_{_V}(r)= \phi_1\phi_2 e^{-r /a},
\end{equation}
where $a$ is a positive constant that represents a characteristic length scale. Using relation (\ref{s2dr}) and Eq. (\ref{S2Debye}), we see the corresponding specific surface $s$ is given by
\begin{equation}
s= \frac{\omega_d d\phi_1\phi_2}{\omega_{d-1}a},
\label{eq:s}
\end{equation} 
which is simply $\pi\phi_1\phi_2/a$ in two dimensions. Note that Debye random media
possess phase-inversion symmetry at the two-point level \cite{torquato2013random}, i.e., $S_2^{(1)}(r;\phi_1,\phi_2)=S_2^{(2)}(r;\phi_2,\phi_1)$. Importantly, we know
that there is a high degeneracy of two-phase media with the same one- and two-point
statistics, but different higher-order correlation functions \cite{torquato2013random, torquato1999exact, jiao2007modeling}. Thus, a model is not uniquely defined only by its two-point correlation function. Debye et al. \cite{debye1957scattering} guessed that the structures corresponding Eq. (\ref{S2Debye}) are those in which one phase consists of ``random shapes and sizes". Two-phase media that realize Eq. (\ref{S2Debye}) for the special case $\phi_1=\phi_2=1/2$ were presented in Ref. \cite{yeong1998reconstructing}. It is also known that certain types of space tessellations in two dimensions have autocovariance functions given by Eq. (\ref{S2Debye}) \cite{St95}. Theoretical analyses indicate that such media are realizable in three and higher dimensions \cite{jiao2007modeling}.\\
\indent In Fig. \ref{fig: reconstruct}, we show select large realizations of Debye random media with phase 2 volume fractions of $\phi_2=0.1$, 0.2, 0.3, 0.4 and 0.5 that we generated using a fast implementation of the Yeong-Torquato construction algorithm (see Sec. IV for details). One can view this algorithm as producing the ``most probable" realizations with an autocovariance function given by Eq. (\ref{S2Debye}). Observe that at small volume fractions, the size of yellow ``islands" varies greatly. As the volume fraction increases, the yellow domains start to connect with each other and percolate at $\phi_2=0.5$. Note that at $\phi_2=0.5$ two phases are not statistically distinguishable, which is in contrast to models of particle dispersions whose phase topologies
are distinctly different from one another
(i.e., do not possess phase-inversion
symmetry), as we show in Sec. VI. In light of phase-inversion symmetry of Debye random media, realizations for $\phi_2=0.6, 0.7, 0.8, 0.9$ are identical to those with $\phi_2=0.4, 0.3, 0.2, 0.1$.

Phase-inversion symmetry implies that the percolation threshold $\phi_2^c$ for Debye random media in $d=2$ is 0.5. To understand this property, we observe that the void phase percolates when $\phi_2<1-\phi_2^c$. Thus, both phases percolate when $\phi_2$ lies in the interval $(\phi_2^c, 1-\phi_2^c)$. However, in two dimensions, two phases cannot percolate in perpendicular directions simultaneously for a finite range of volume fractions. Thus, it is reasonable to argue that the interval will shrink to a single point, i.e., $\phi_2^c=0.5$. A systematic study of the percolation behavior of Debye random media requires not only the ability to generate large samples, but also a large number of them. Although we solve the former problem in the next section, generating a large number of realizations is still computationally challenging and so the percolation properties will not be studied in this paper. However, visual inspection of the realizations shown in Fig. \ref{fig: reconstruct} are consistent with the percolation threshold occurring at $\phi_2=0.5$.      

\section{Accelerated Yeong-Torquato Construction Algorithm}  

To construct realizations of Debye random media, we apply a variation of the (re)construction algorithm formulated by Yeong and Torquato \cite{yeong1998reconstructing}, which has been applied by a variety of different investigators \cite{jiao2007modeling,jiao2009superior, chen2015dynamic, karsanina2018hierarchical,vcapek2018importance,li2018transfer,pant2015multigrid,gerke2019calculation}. The procedure treats the (re)construction problem as an energy-minimization problem and solves it by simulated annealing. Consider constructing a digitized two-phase system contained within a hypercube in $d$ dimensions of side length $L$ and $N^d$ pixels (voxels), which is subjected to periodic boundary conditions. A fictitious energy $E$ is defined as the squared differences between the target and simulated correlation functions. Then pairs of pixels from different phases are swapped according to the Metropolis rule. In this paper, our target is the two-point correlation function $S_2$ given in Eq. (\ref{S2Debye}), the energy is given by 
\begin{equation} \label{energy}
E=\sum_i \left[\chi_{_V}(r_i)-\hat{\chi}_{_V}(r_i)\right]^2,
\end{equation}
where $\chi_{_V}(r_i)$ and $\hat{\chi}_{_V}(r_i)$ are simulated and target autocovariance functions (note that this method was designed to target multiple statistical descriptors \cite{yeong1998reconstructing}). Here $r_i$ runs over all the distances formed by pairs of pixels of the target phase. In contrast to the orthogonal sampling method used by Yeong and Torquato \cite{yeong1998reconstructing}, we sample 
two-point statistics in all directions, as described in Ref. \cite{jiao2007modeling}. Note that by using simulated annealing we are effectively sampling ``entropically favored" or equilibrium realizations subject to the energy form given in Eq. (\ref{energy}). Since only the volume fraction and two-point correlation function are constrained, this procedure implements the maximal entropy principle. Thus, we argue that our procedure produces ``most probable" realizations.
\begin{figure}[h]
\centering
\includegraphics[width=8cm, height=6cm, clip=true]{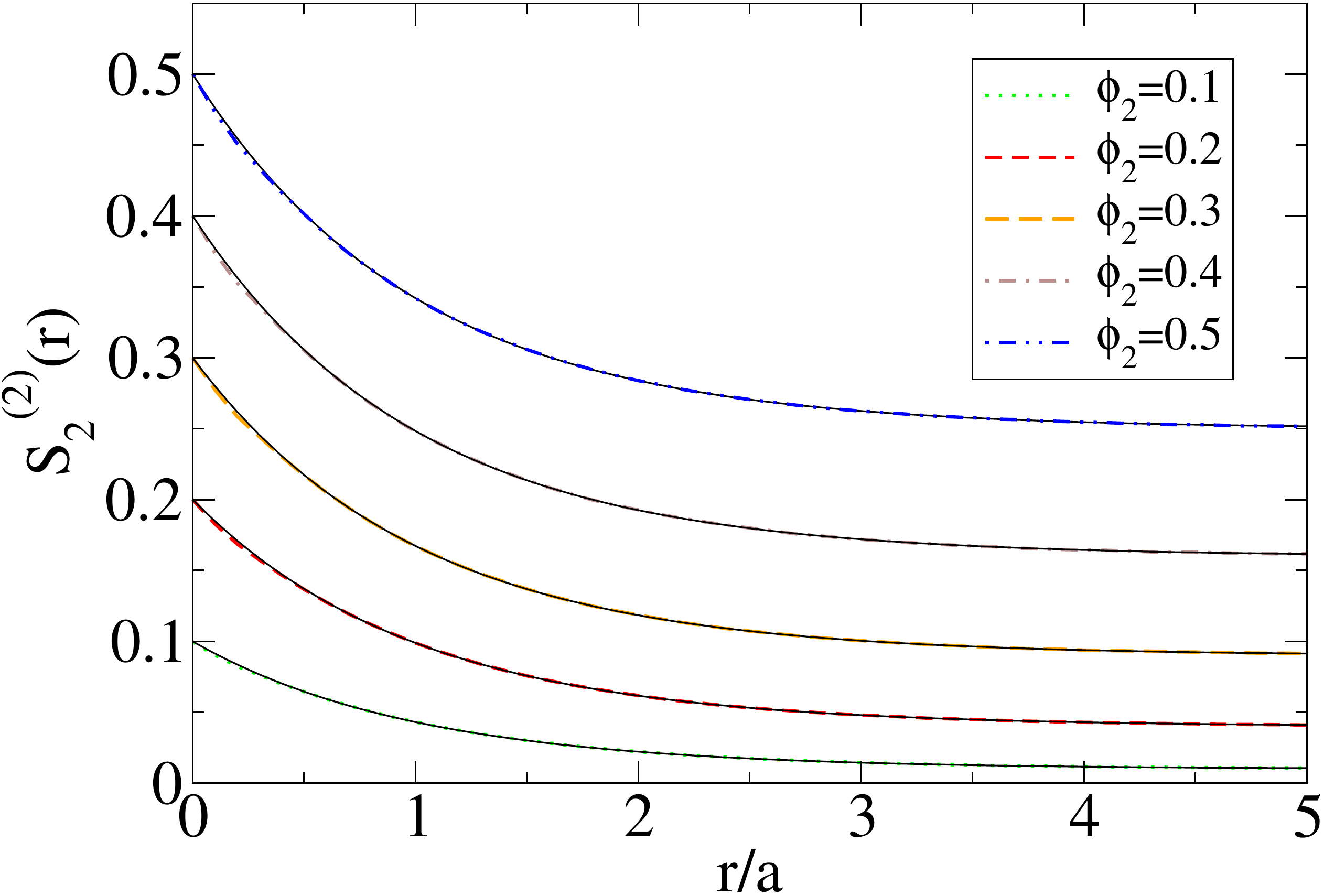}
\caption{Comparison between the simulated $S_2(r)$'s of the constructed Debye random media (shown as dashed curves) and
the targeted ones (shown as solid black curves) for different volume
fractions of phase 2: Curves from top to bottom span from $\phi_2=0.5$
to $\phi_2=0.1$ in increments of 0.1. The simulated two-point functions are in excellent
agreement with the corresponding target functions.}
\label{fig: constructs2}
\end{figure}      

It has been suggested that Debye random media are expected to have relatively large ``holes" compared to other systems, such as overlapping spheres, due to the fact that the autocovariance function of Debye random media has infinite support, whereas the one for overlapping spheres has finite support \cite{torquato2020predicting}. This conjectured large-hole property of Debye random media implies that an accurate characterization of their microstructures demands an ability to systematically construct sufficiently large samples (e.g., digitized systems consisting $500^d$ voxels or larger in $d$ dimensions) in order to start to observe holes of large size. However, applying the general Yeong-Torquato technique can be challenging for such tasks. It is expected that the number of Monte Carlo steps required grows at least as fast as $\mathcal{O}(N^d)$, since one wants to ensure that on average each pixel (voxel) will be swapped for a sufficient number of times. For each swap, when updating $S_2(r)$, we need to consider every pair of pixels (voxels) formed by the chosen pixels (voxels) and the rest of them. This again leads to $\mathcal{O}(N^d)$ operations. Thus the complexity of the entire algorithm scales as $\mathcal{O}(N^{2d})$. This scaling behavior is clearly too demanding even for two dimensions, and almost impractical for three dimensions. 

To tackle this challenge, we tailor the Yeong-Torquato
construction procedure to appreciably speed-up sampling the autocovariance function for Debye random media and related ones. Specifically, we apply a cutoff of the two-point correlation function at a length scale $l_c$ that is much larger than the characteristic length scale $a$ but smaller than the system size. For the large samples we are interested here, the system size $L$ is much greater than the characteristic length $a$. As a result, the autocovariance function given in Eq. (\ref{S2Debye}) is essentially zero for distance $r$ much larger than $a$. If the long-range behavior ($r\geq l_c$) of the system is not relevant, then updating $S_2(r)$ for those pixel (voxel) pairs can be very inefficient. By choosing a cutoff $l_c$ that is still much larger than $a$ but smaller than the system size $L$, we can reduce the complexity of updating $S_2(r)$ to $\mathcal{O}(l_c^d)$ without sacrificing the accuracy of the construction. This trick brings the complexity of the entire algorithm down to $\mathcal{O}(N^d)$ and greatly reduces the computing time. Note that this efficient method can also be applied to model other disordered structures whose autocovariance function decays fast enough so that it is essentially zero beyond this {\it correlation length}.  

We also apply a final refining process after a fraction of the total Monte Carlo steps to eliminate small isolated ``islands" (pixels or voxels) of one phase in a ``sea" of the other phase that should be present. To do so, we keep track of the list of pixels (voxels) that are on the interfaces between two phases and only select from this group. Each update will only change the list locally, so the extra computing time compared to a standard Monte Carlo step is very minor, if implemented accordingly.   

To systematically construct Debye random media, we also purposely tune parameters such that they change with volume fractions/sizes automatically. Specifically, we choose the initial temperature $\sim \phi_1\phi_2/N^d$ and the number of Monte Carlo steps $\sim \phi_1\phi_2N^d$. Note that these choices also explicitly make our constructions phase-inversion symmetric, since the code for constructing a sample with volume fraction $\phi_2=\phi$ is exactly the same as the one for constructing a sample with $\phi_2=1-\phi$. 

Since it is still a considerably challenging computational task to construct large Debye random media in three dimensions, we focus here on generating such media in two dimensions. Nonetheless, our two-dimensional (2D) results have interesting implications in higher dimensions, as we will discuss in Sec. VII. In simulations, we choose $a=5$ and $l_c=10a$, and construct samples with size $L=501$ at volume fractions $\phi_2=0.1, 0.2, 0.3, 0.4, 0.5$. The simulated $S_2(r)$ (shown in dashed lines) for each volume fractions compared with their targets (shown in solid black lines) are plotted in Fig. \ref{fig: constructs2}, one can see that they match extremely well. Specifically, we measure how well the construction is by the average of the absolute values of discrepancies of the two-point correlation function \cite{jiao2008modeling}, defined as $\Delta S_2=1/N_L\sum_{r}|\delta S_2(r)|$, where $N_L$ is the number of bins. We find the average discrepancies $\Delta S_2$ for constructed samples are quite small as $0.6\sim2\times 10^{-4}$, which are even smaller than those reported in Ref. \cite{jiao2008modeling}. Representative digitized images of 2D Debye random media obtained from the construction algorithm at different volume fractions are shown in Fig. \ref{fig: reconstruct}.

\section{Results for Other Statistical Descriptors}
In this section, we compute the other aforementioned statistical descriptors for our constructed Debye random media. Specifically, we compute the two-point surface correlation functions $F_{\mathrm{ss}}(r)$ and $F_{\mathrm{sv}}(r)$, pore-size probability density function $P(\delta)$, lineal-path function $L(z)$ for the matrix phase, and matrix chord-length probability density function $p(z)$. 

For simpler models of two-phase media consisting of spheres in a matrix, the determination
of such microstructural descriptors can be explicitly represented as an infinite series that generally requires
an infinite amount of information via the $n$-particle correlation functions
$g_1, g_2, g_3, \ldots$ \cite{To86i}. Since the $g_n$
are only known exactly for uncorrelated spheres (overlapping spheres), one must generally devise
approximation formulas for the descriptors. In the case
of Debye random media, such explicit representations are not known
and so it is desirable to obtain semi-analytical or empirical formulas for the relevant descriptors. Importantly, we find simple and accurate semi-analytical expressions for $F_{\mathrm{ss}}(r)$ and $F_{\mathrm{sv}}(r)$. We also present an empirical fitting function for the pore-size probability density function $P(\delta)$. We find that in the case of the lineal-path function and chord-length density function are well-approximated by the corresponding functional forms for a system of overlapping disks of radius nearly equal to the characteristic length scale $a$ [cf. Eq. (\ref{S2Debye})]. All results are averaged over 10 different constructions.

\begin{figure*}[]
\centering
\subfigure[]{
\includegraphics[clip=true, width=5cm, height=3.5cm]{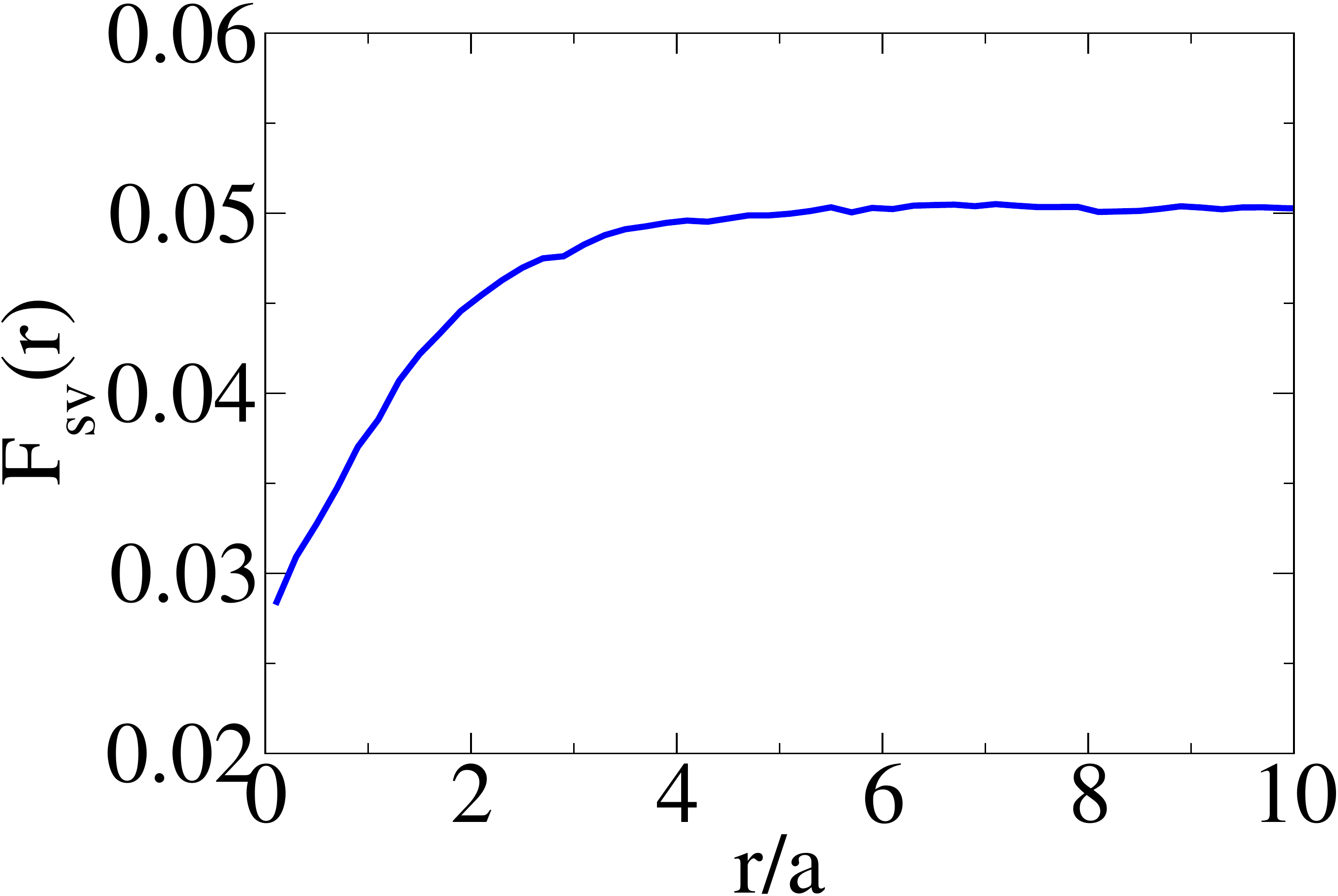}
}
\subfigure[]{
\includegraphics[clip=true, width=5cm, height=3.5cm]{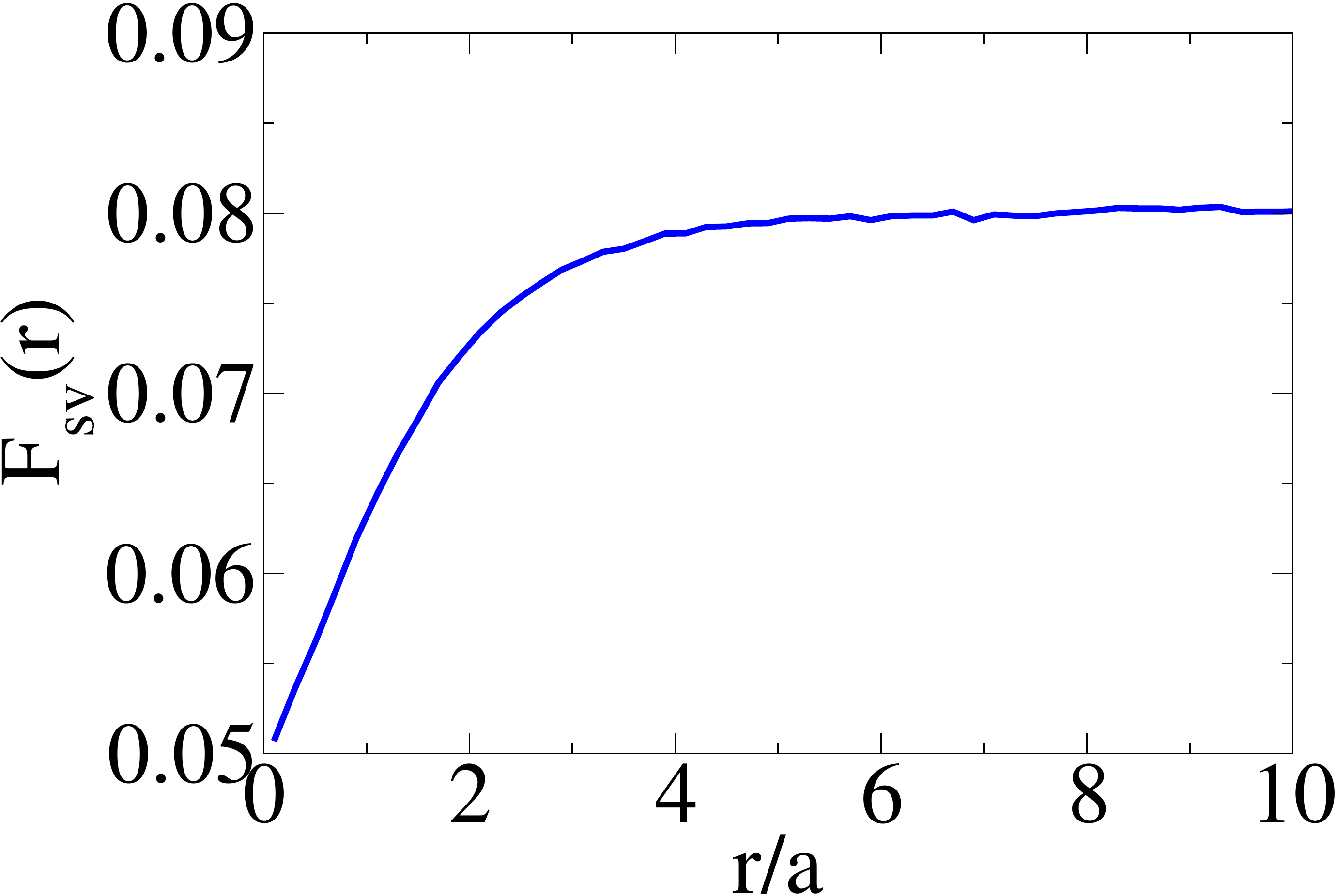}
}
\subfigure[]{
\includegraphics[clip=true, width=5cm, height=3.5cm]{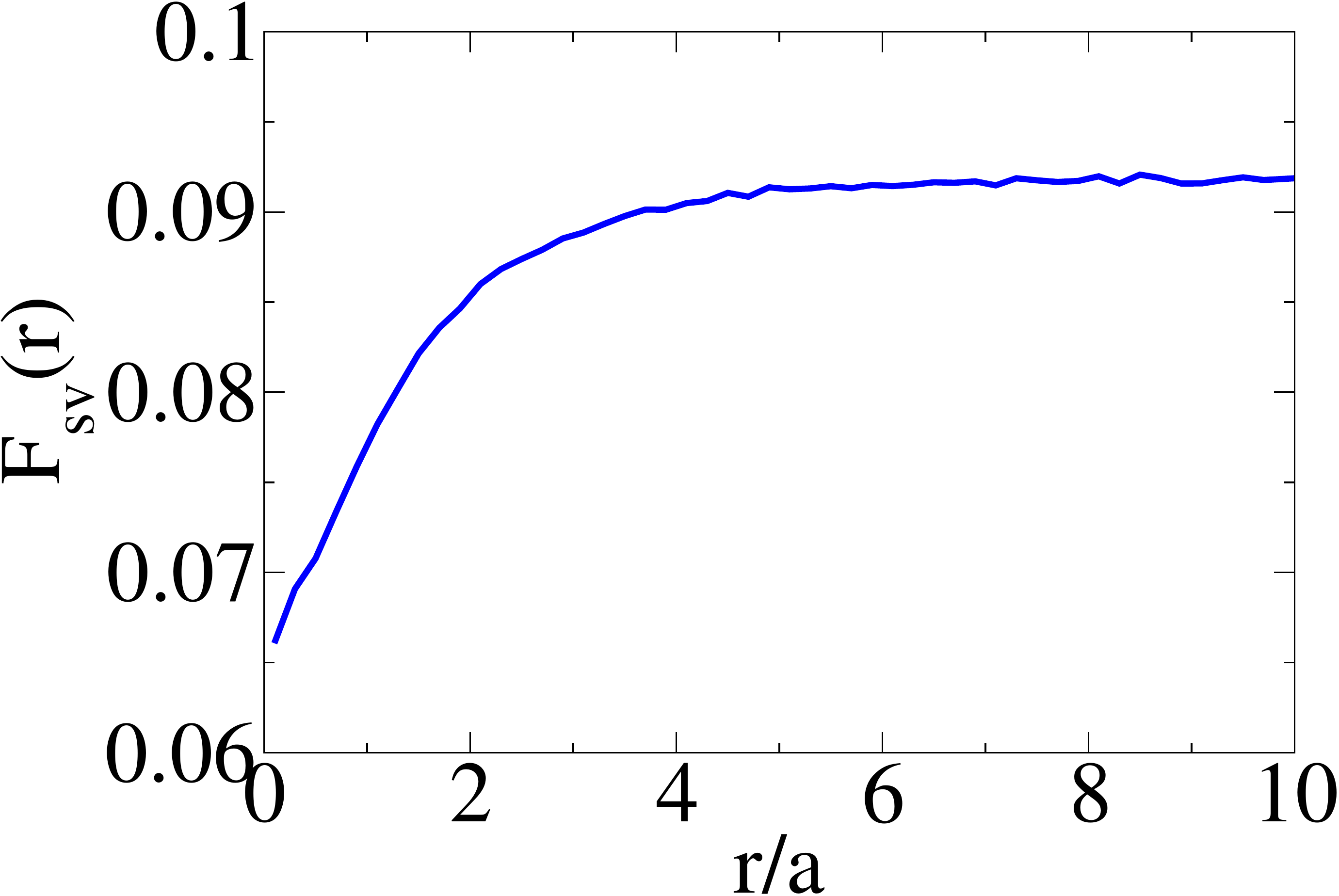}
}
\subfigure[]{
\includegraphics[clip=true, width=5cm, height=3.5cm]{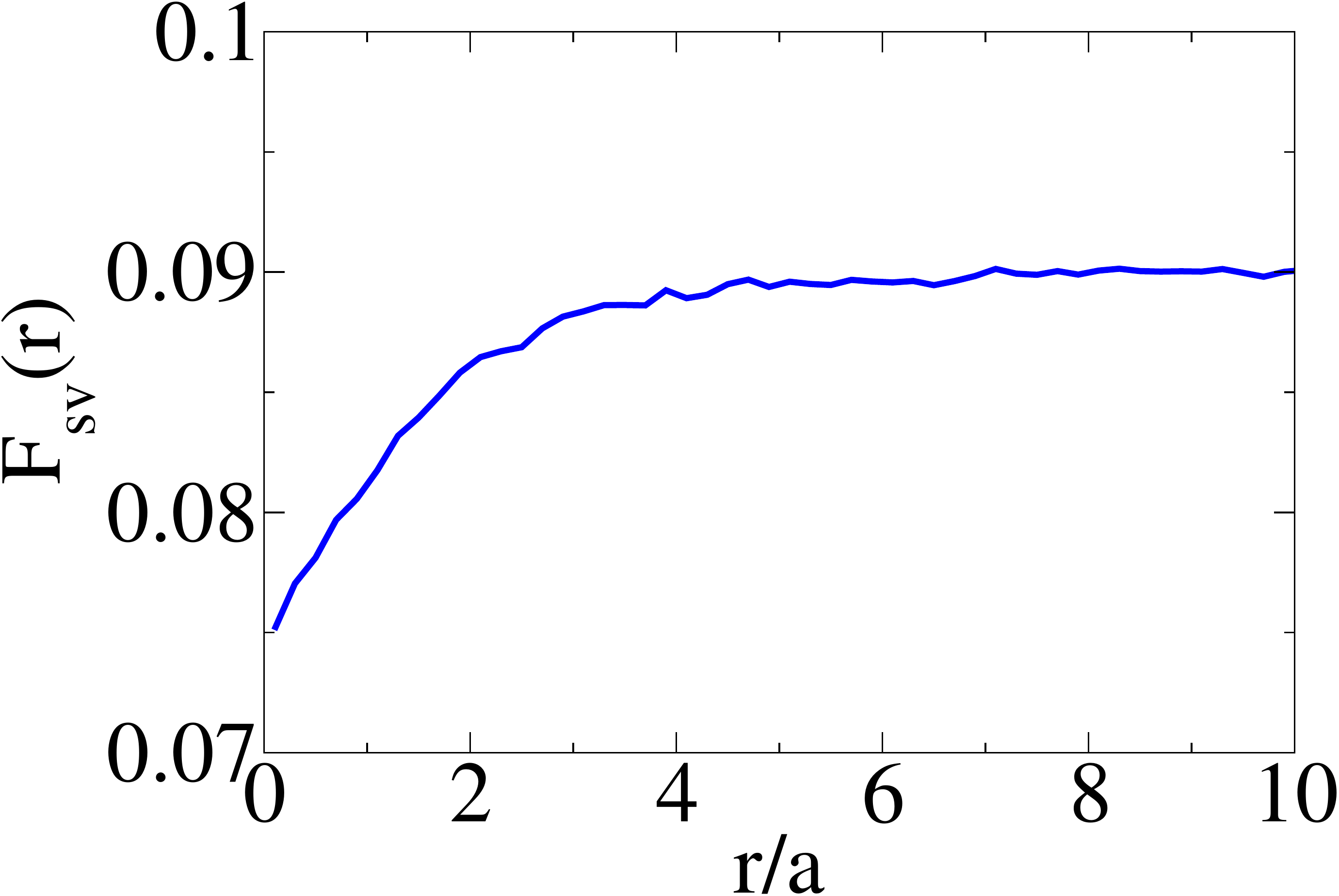}
}
\subfigure[]{
\includegraphics[clip=true, width=5cm, height=3.5cm]{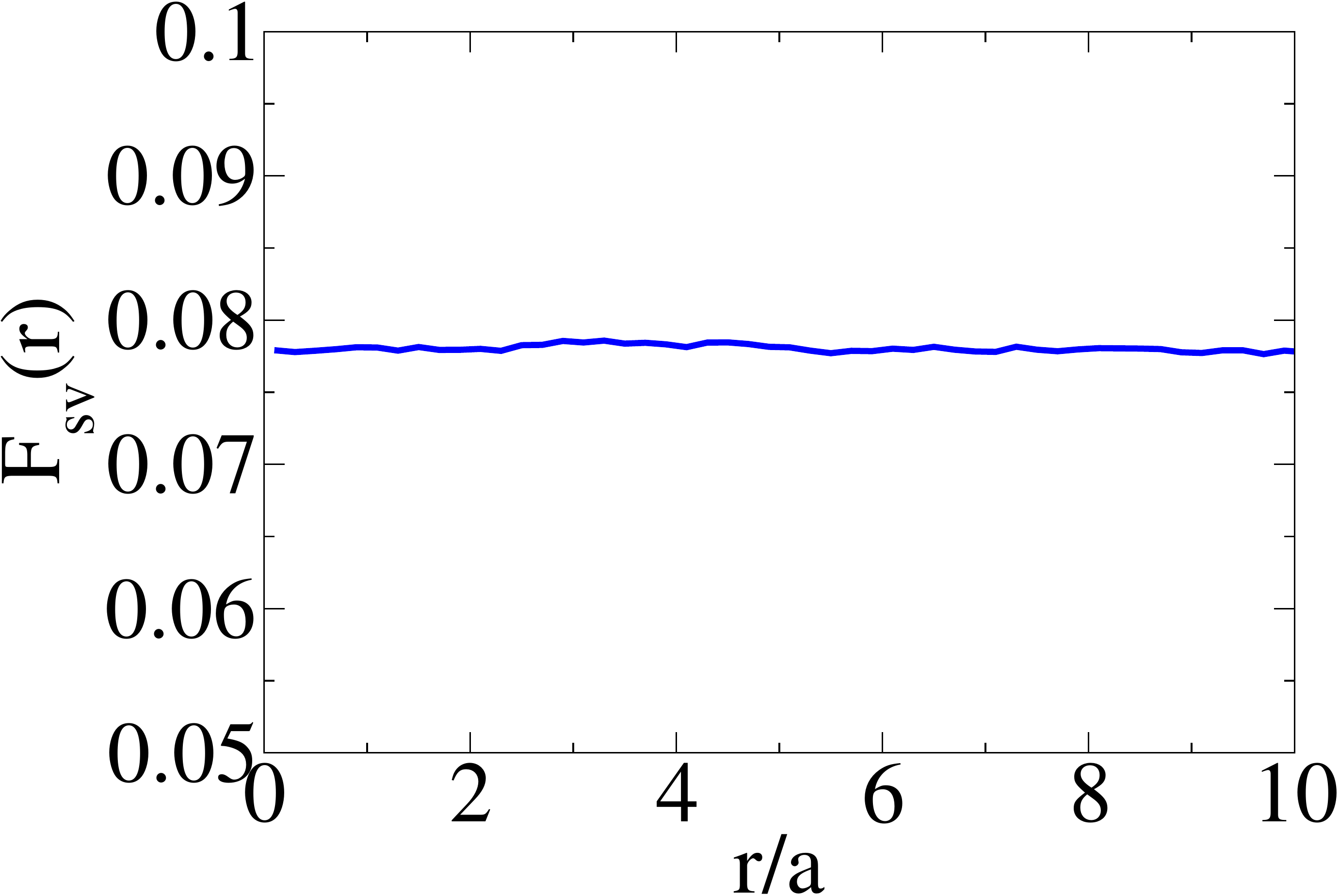}
}
\caption{Surface-void correlation function $F_{\mathrm{sv}}(r)$ for Debye random media at different volume fractions. (a) $\phi_2=0.1$. (b) $\phi_2=0.2$. (c) $\phi_2=0.3$. (d) $\phi_2=0.4$. (e) $\phi_2=0.5$.}
\label{fig: Fsv}
\end{figure*}

\subsection{Surface Correlation Functions}
\begin{figure*}[]
\centering
\subfigure[]{
\includegraphics[clip=true, width=5cm, height=3.5cm]{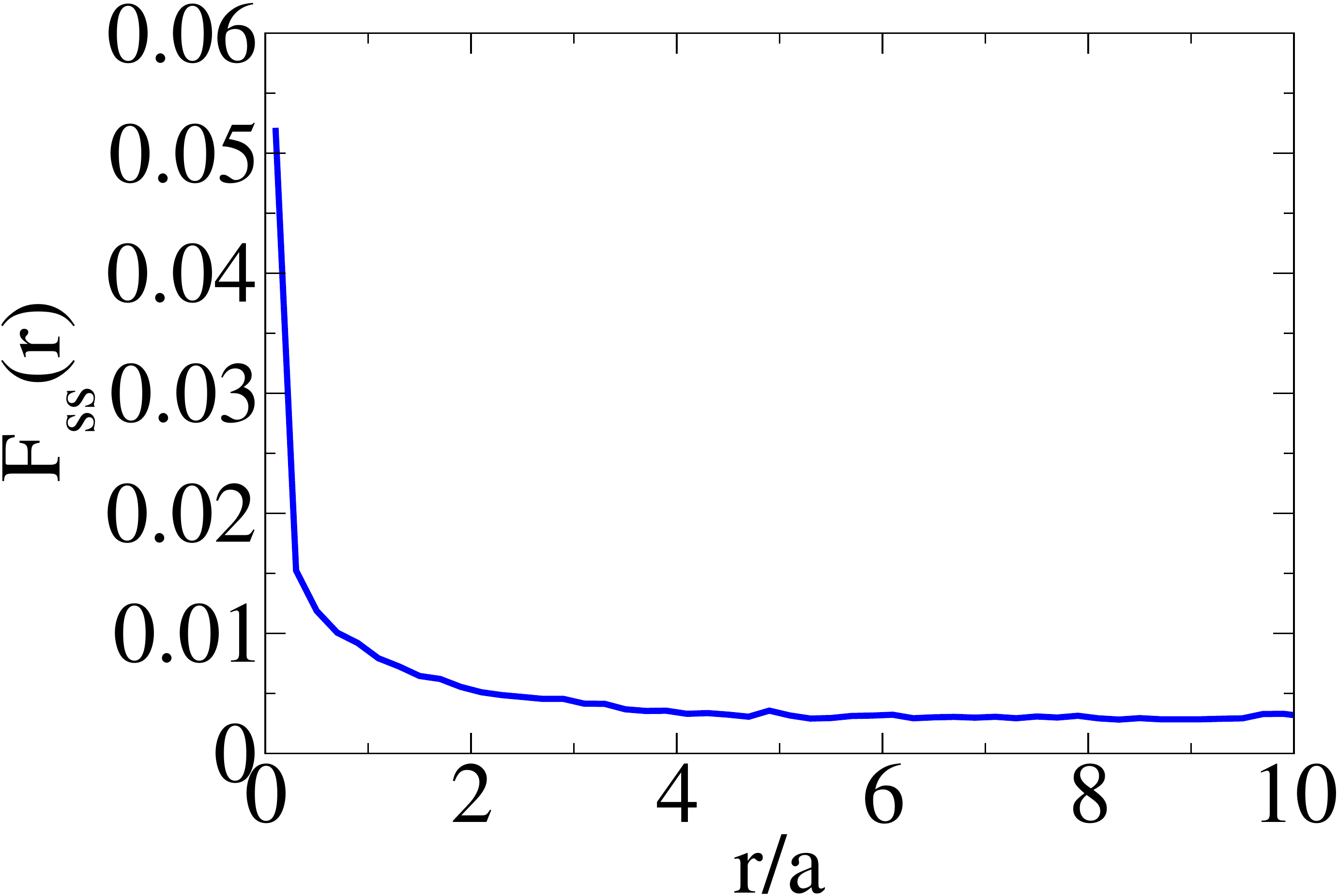}
}
\subfigure[]{
\includegraphics[clip=true, width=5cm, height=3.5cm]{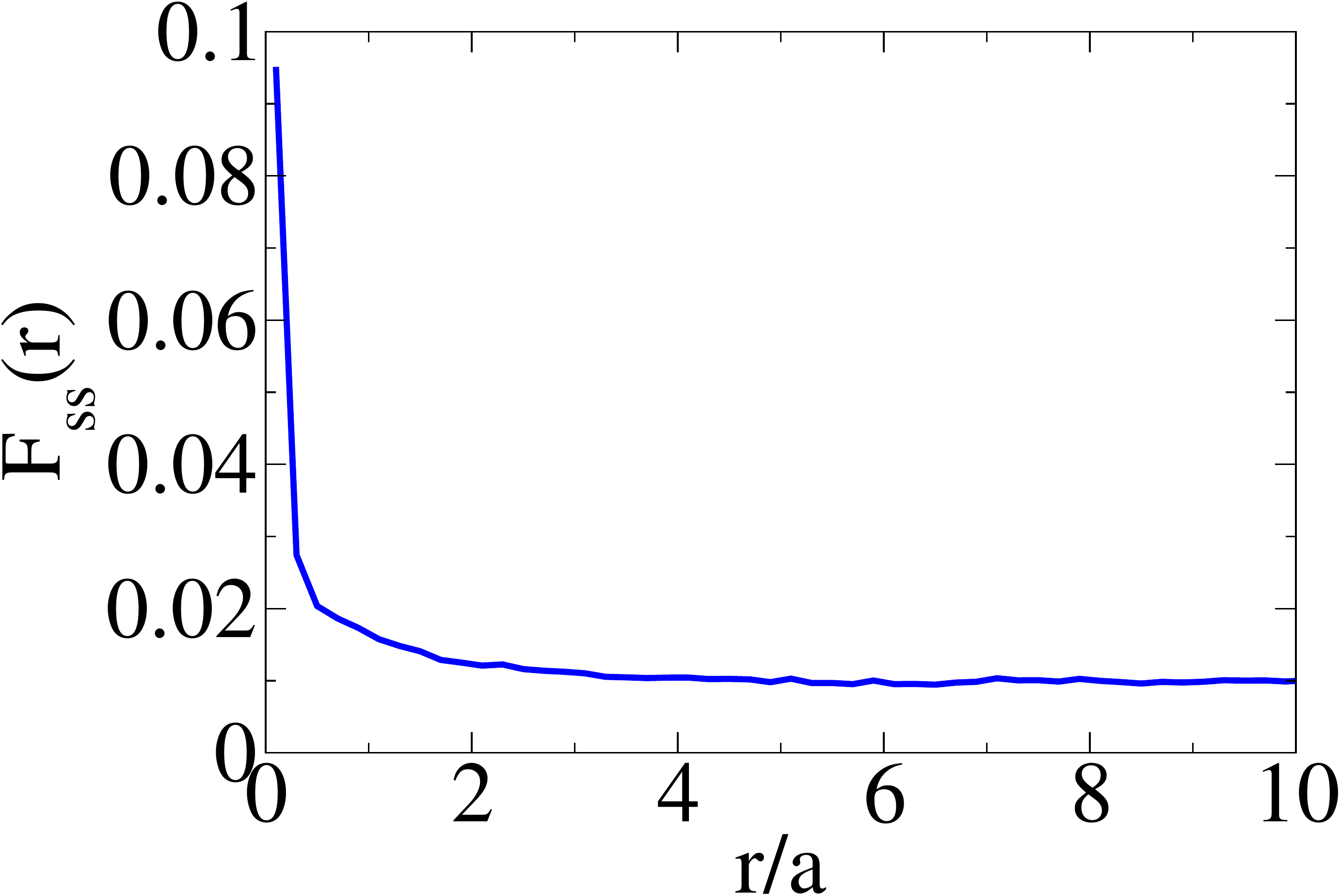}
}
\subfigure[]{
\includegraphics[clip=true, width=5cm, height=3.5cm]{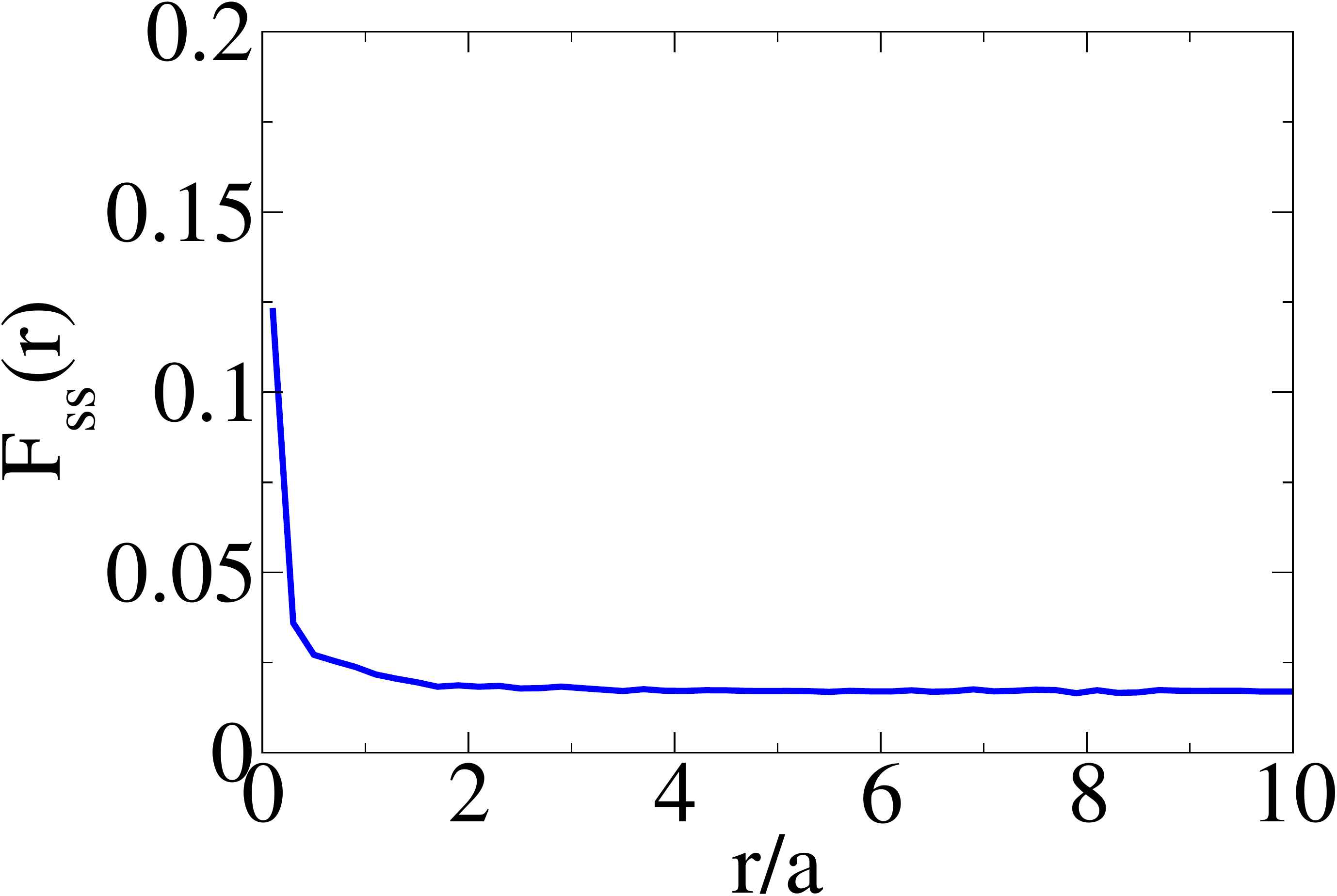}
}
\subfigure[]{
\includegraphics[clip=true, width=5cm, height=3.5cm]{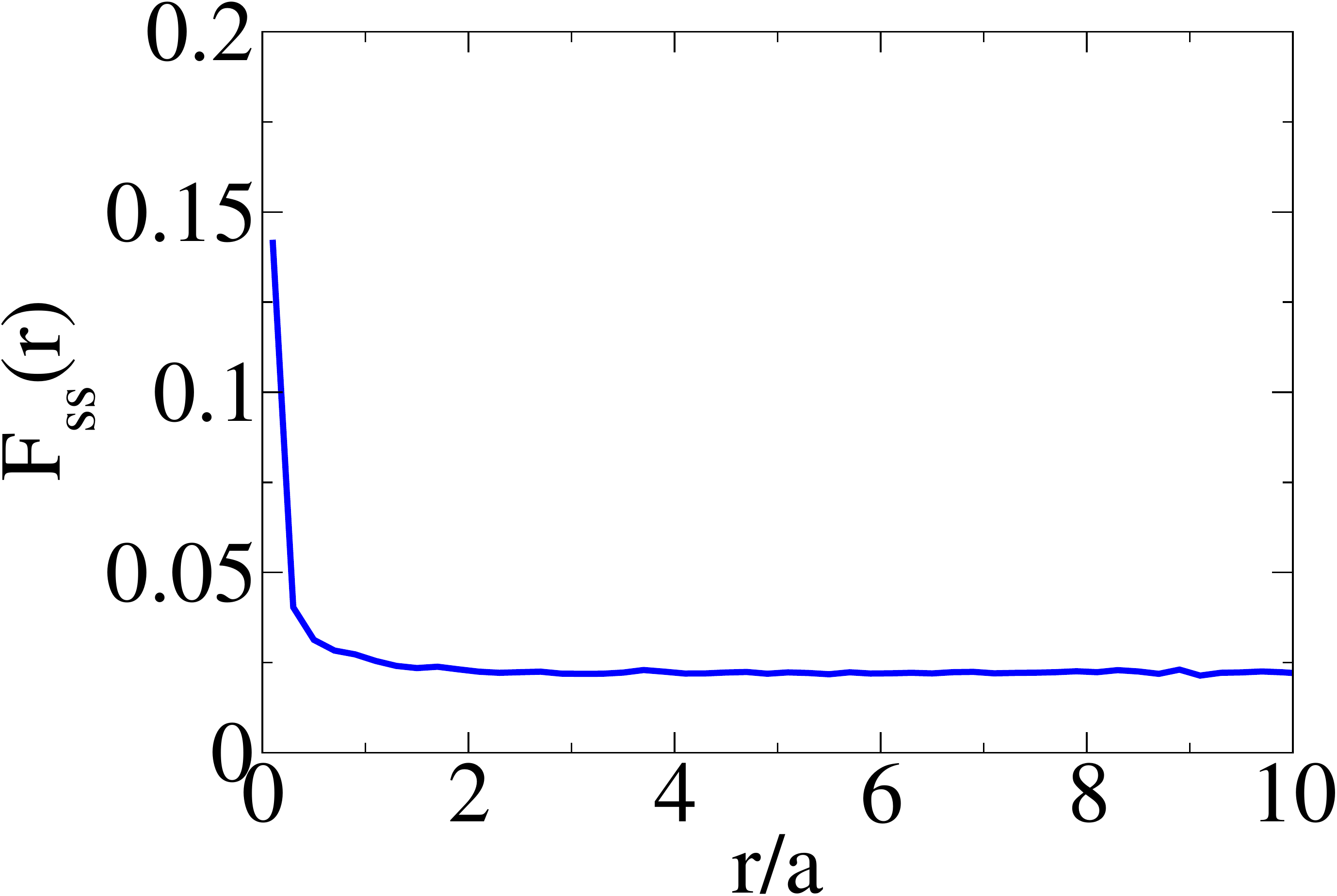}
}
\subfigure[]{
\includegraphics[clip=true, width=5cm, height=3.5cm]{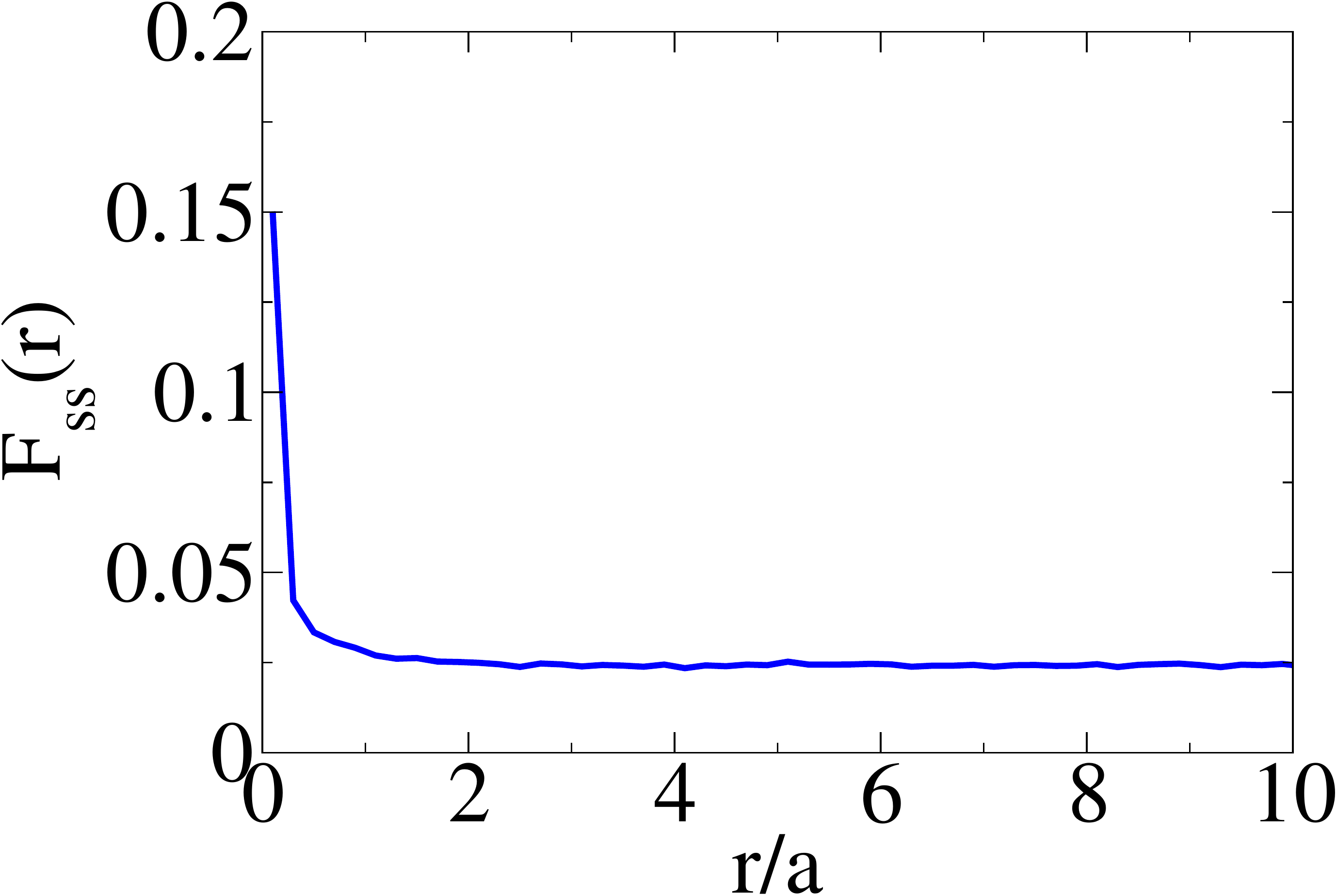}
}
\caption{Surface-surface correlation function $F_{\mathrm{ss}}(r)$ for Debye random media at different volume fractions. (a) $\phi_2=0.1$. (b) $\phi_2=0.2$. (c) $\phi_2=0.3$. (d) $\phi_2=0.4$. (e) $\phi_2=0.5$.}
\label{fig: Fss}
\end{figure*}

The two-point correlation function $S_2(r)$ for Debye random media is already known by definition. Thus, here we study other two-point correlation functions, namely the surface-surface correlation function $F_{\mathrm{ss}}(r)$ and surface-void correlation function $F_{\mathrm{sv}}(r)$. The efficient calculation of these correlation functions for general two-phase systems has been made possible recently due to algorithms developed by Ma and Torquato \cite{ma2018precise}. We apply such algorithms to the constructed samples and average the results for 10 realizations at each volume fraction. Specifically, in the step of converting the two-phase media into scalar fields, we choose thresholds such that the volume fraction is kept the same. We also tune the Gaussian filters such that the specific surface $s$ is consistent with the theoretical value. The results for $F_{\mathrm{sv}}(r)$ and $F_{\mathrm{ss}}(r)$ are shown in Figs. \ref{fig: Fsv} and \ref{fig: Fss}, respectively. 

The behavior of $F_{\mathrm{ss}}(r)$ is dominated by the divergent behavior at small $r$, so we focus on $F_{\mathrm{sv}}(r)$ first. Clearly, the curves in Fig. \ref{fig: Fsv} are sigmoid-like. This motivates us to fit the computed $F_{\mathrm{sv}}(r)$ by logistic functions. Interestingly, the results strongly suggest that $F_{\mathrm{sv}}(r)$ can be described by a simple analytical form:
\begin{equation} \label{eq:debyeFsv}
\begin{split}
F_{\mathrm{sv}}(r) &= -\frac{2(F_{\mathrm{sv}}(r=\infty)-F_{\mathrm{sv}}(0))}{\exp(r/a)+1}+F_{\mathrm{sv}}(r=\infty)\\
&=\frac{1-\phi_2+\phi_2\exp(-r/a)}{1+\exp(-r/a)}s,
\end{split}
\end{equation}
where $s$ is given by Eq. (\ref{eq:s}). This function fully recovers the exact results at $r=0$ and $r=\infty$. Note that the origin is exactly at the midpoint of the logistic function.  

Notice that the two-point correlation function for the void phase is $S_2^{(1)}=\phi_1\phi_2\exp(-r/a)+\phi_1^2$. We can rewrite Eq. (\ref{eq:debyeFsv}) as:
\begin{equation}
F_{\mathrm{sv}}(r)=\frac{\pi\phi_2}{a}\frac{1}{1+\exp(-r/a)}S_2^{(1)}(r).
\end{equation}
This functional form is reminiscent of that of the surface-void correlation function for overlapping spheres, the expression in two dimensions is derived and shown in Sec. VI. A, see Eq. (\ref{eq:2dfsv}).  
Compare the expressions we immediately see that the sharp transition at $r=2R$ for overlapping disks is replaced by a smooth decaying function in the case of Debye random media. This observation is consistent with the argument that Debye random media consist of domains of ``random shape and size". Moreover, using the relation in Eq. (\ref{fsvdr2}), we find the specific Euler characteristic is $\pi(\phi_1-\phi_2)\phi_1\phi_2/4a^2$. It is clear that it vanishes as $\phi_2 \rightarrow 0.5$, as percolation happens.

One may be tempted to arrive at the effective form of $F_{\mathrm{ss}}(r)$ using the same reasoning, but this is not tenable. We know that for Debye random media $F_{\mathrm{ss}}(r)$ must be invariant under the transform $\phi_2 \rightarrow (1-\phi_2)$. However, the functional form of $F_{\mathrm{ss}}(r)$ for overlapping disks (see Eq. (\ref{eq:2dfss})) 
does not have the property since $S_2(r)$ is clearly not invariant under the transform. Instead, after many tries, we find that the following form fits our data excellently for all volume fractions:  
\begin{equation} 
\begin{split}
F_{\mathrm{ss}}(r)=\frac{\pi^2}{a^2}\phi_1^2\phi_2^2+\frac{1}{ar}\phi_1\phi_2\exp(-r/a)\\+\frac{1}{2a^2}\frac{\exp(-r/a)}{1+\exp(-r/a)}|\phi_2-\phi_1|
\end{split}
\label{eq:debyeFss}
\end{equation}
This function also fully recovers the exact results at small and large $r$ limits and satisfies the phase-interchange invariance.

\subsection{Pore-Size Functions}  
\begin{figure}[]
\centering
\subfigure[]{
\includegraphics[width=8cm, height=5cm]{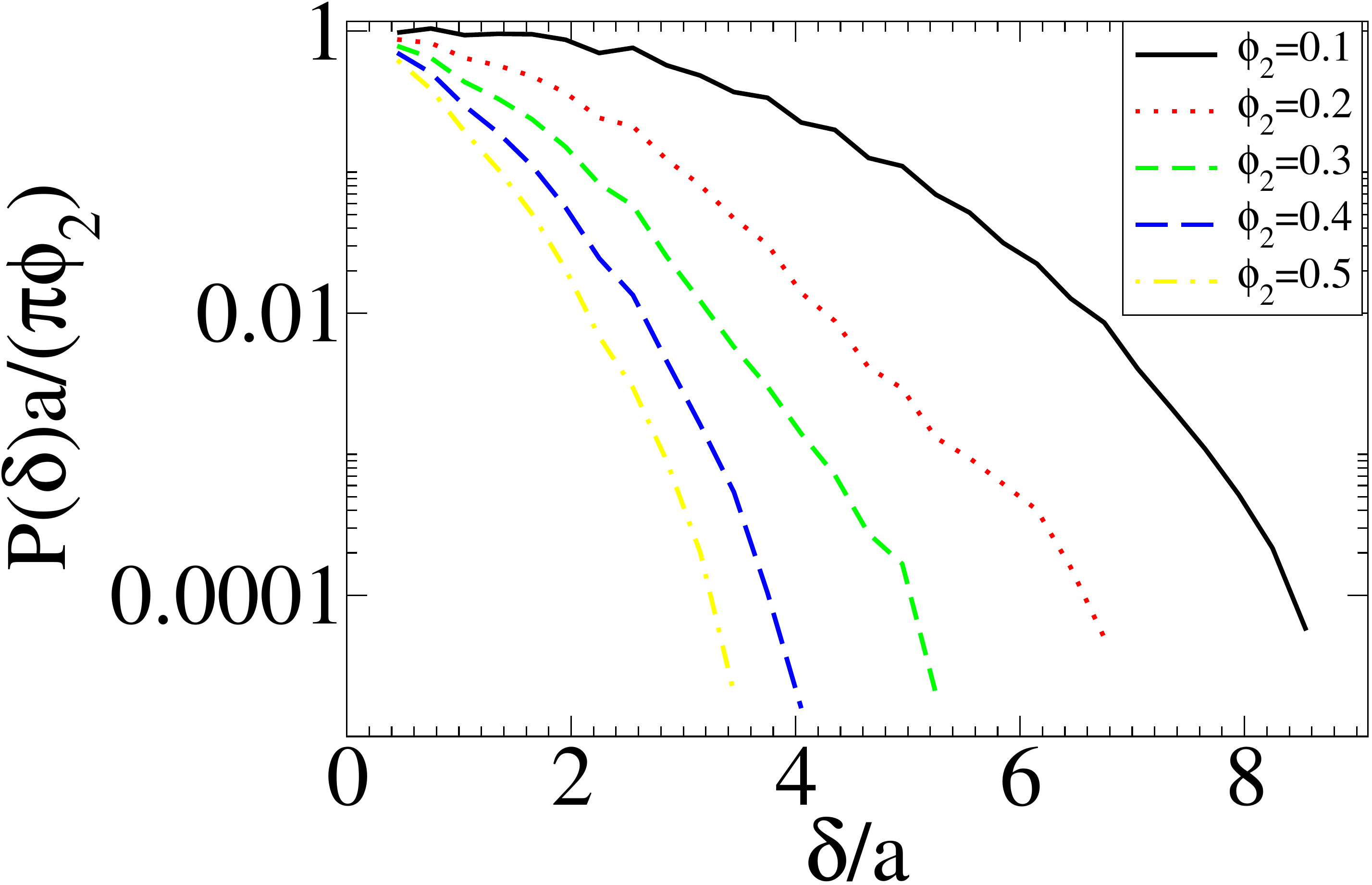}
}
\subfigure[]{
\includegraphics[width=7cm, height=5cm]{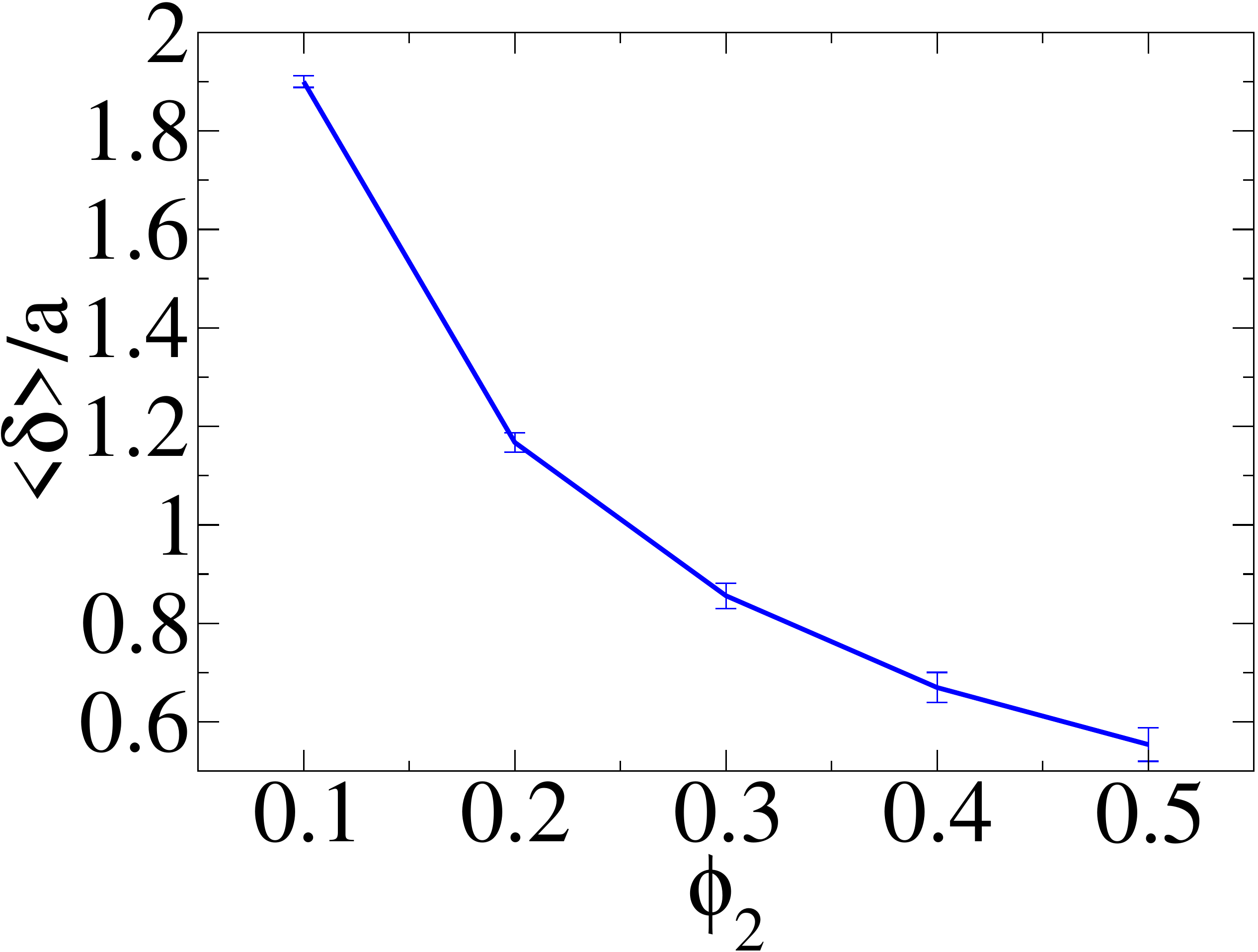}
}
\caption{(a) The pore-size probability density function $P(\delta)$ for Debye random media for selected volume fractions $\phi_2=0.1, 0.2, 0.3, 0.4$ and 0.5. (b) The mean pore size $\langle \delta \rangle$ as a function of volume fraction.}
\label{fig:pore}
\end{figure}

In Fig. \ref{fig:pore} we show the computed pore-size probability density function $P(\delta)$ and the associate mean pore size $\langle \delta \rangle$ for different volume fractions. Here $P(\delta)$ is scaled by $\phi_1/s$ to bring its value at the origin to unity. Interestingly, these results indeed imply that Debye random media possess large ``holes", as shown in the next section by comparing the results to those of other models. Guided by the scaled-particle theory \cite{torquato2013random, torquato1995nearest} used to derive $P(\delta)$ for equilibrium hard spheres [see Eq. (\ref{eq:hardpore})], we propose the following pore-size probability density function: 
\begin{equation}
P(\delta)=(p_0+2p_1\delta)\exp(-p_1\delta^2-p_2\delta),
\label{eq:porefit}
\end{equation}
where $p_0$, $p_1$ and $p_2$ are coefficients. First, note that $P(\delta=0)=s/\phi_1$ enables us to determine $p_0=\pi\phi_2/a$. Interestingly, we find that by setting $p_2=\pi\phi_2/a$ (this is also consistent with what we obtain if $p_2$ is fitted as a free parameter), the normalization condition $\int_{0}^{\infty}P(\delta) d \delta=1$ is always satisfied. Fitting this functional form (\ref{eq:porefit}) for $P(\delta)$ (with $p_1$ as the only free parameter) to the data, we find the following approximation $p_1=(1.05\phi_2-2.41\phi_2^2+4.16\phi_2^3)/a^2$. Using this empirical equation, we can analytically compute the mean pore size $\langle \delta \rangle$, which is given by
\begin{equation}
\begin{split}
\frac{\langle \delta \rangle}{a}=&\frac{\sqrt{\pi}e^{\pi^2\phi_2^2/4(1.05\phi_2-2.41\phi_2^2+4.16\phi_2^3)}}{2\sqrt{1.05\phi_2-2.41\phi_2^2+4.16\phi_2^3}} \\
&(1-\erf(\frac{\pi\phi_2}{2\sqrt{1.05\phi_2-2.41\phi_2^2+4.16\phi_2^3}})),
\end{split}
\label{eq:meanporefit}
\end{equation}
where $\erf(x)$ is the error function. The relative error of the mean pore size computed from this expression compared to the simulated values varies from $4\%$ to $7\%$ for the considered volume fractions. 

The empirical formula (\ref{eq:meanporefit}) for the mean pore size appears to qualitatively capture 
the appropriate asymptotic behavior in the limits $\phi_2 \rightarrow 0$ and $\phi_2 \rightarrow 1$, even though it was not obtained using information at these extreme limits. When $\phi_2 \rightarrow 0$, formula (\ref{eq:meanporefit}) yields $\langle \delta \rangle/a \sim \phi_2^{-1/2}$. This scaling can be physically explained by the fact that the solid phase, in this dilute limit, can be regarded to consist of ``islands" of effective size $\sim a$. Moreover, the average ``density" of these islands would be approximately $\phi_2/a^2$. Thus, the typical distance between two islands is given by $a\phi_2^{-1/2}$, which is proportional to the mean pore size. On the other hand, as $\phi_2 \rightarrow 1$, the pore space will consist of islands of size $\sim a$, which is consistent with the prediction of (\ref{eq:meanporefit}) that $\langle \delta \rangle \sim a$.      

\subsection{Lineal-Path Function}  
\begin{figure}[]
\centering
\subfigure[]{
\includegraphics[width=8cm, height=5cm]{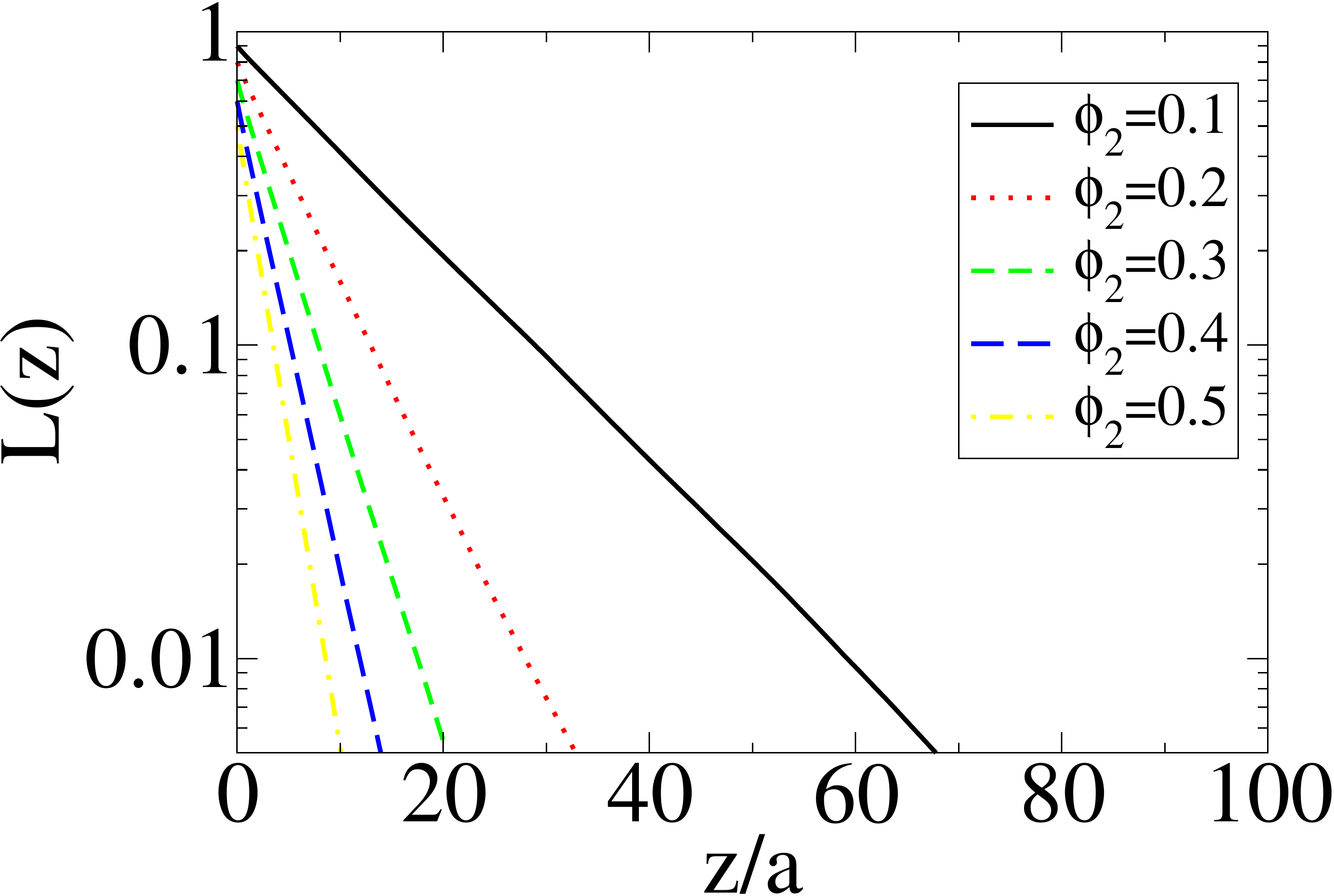}
}
\subfigure[]{
\includegraphics[width=7cm, height=5cm]{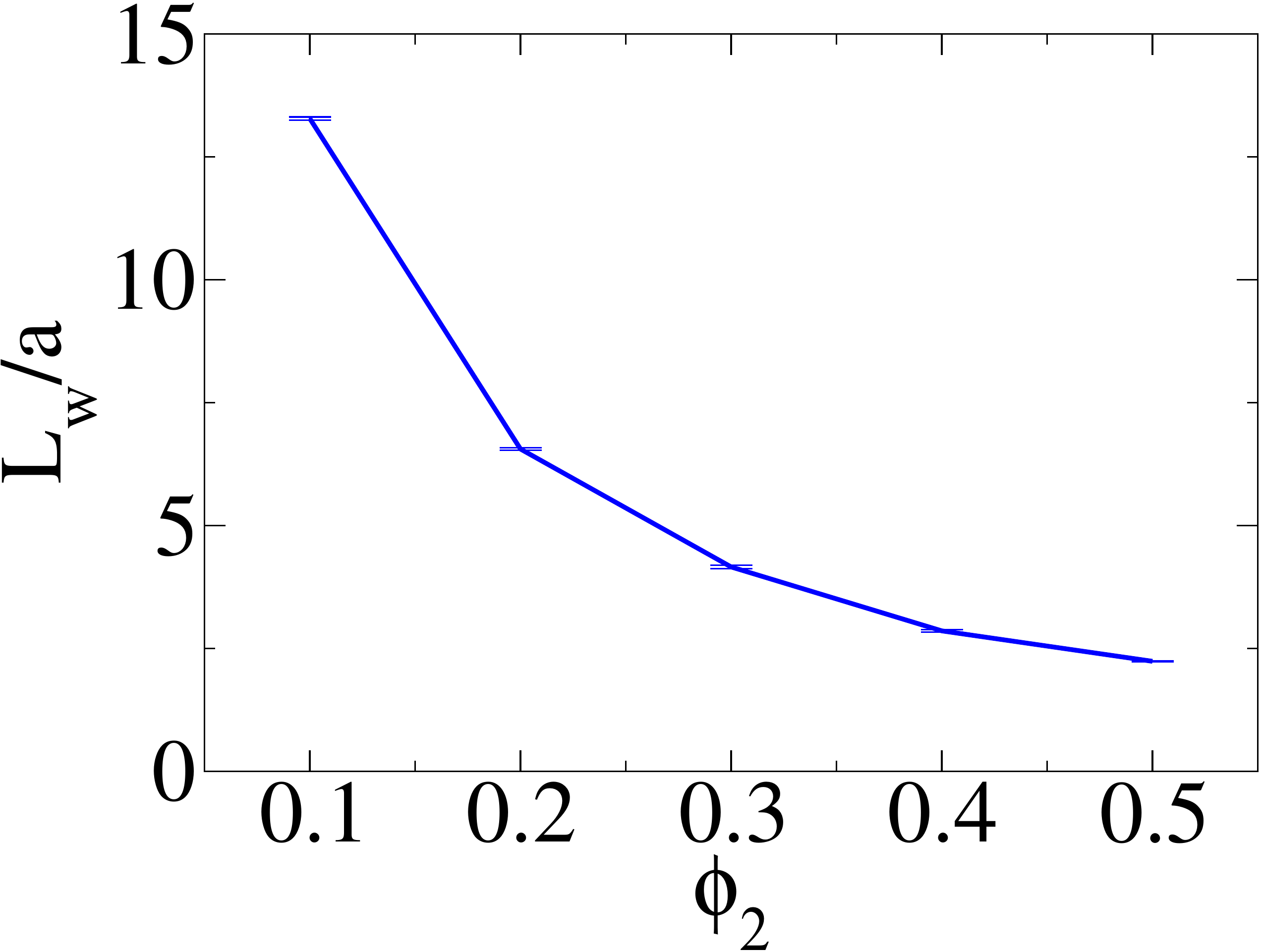}
}
\caption{(a) The lineal-path function $L(z)$ for Debye random media for different volume fractions of phase 2, $\phi_2$. (b) The fitted $L_w$ as a function of $\phi_2$.}
\label{fig:lp}
\end{figure}

In Fig. \ref{fig:lp}(a) we show the lineal-path function for different volume fractions. It is obvious that in all cases the lineal-path function exhibits an exponential decay behavior, where as the volume fraction of the pore space shrinks it decays more rapidly. We can write $L(z)$ as 
\begin{equation}
L(z)=\phi_1\exp(-z/L_w), 
\label{eq:lpfit}
\end{equation}
where $L_w$ can be understood as an average {\it lineal} size of the pore domain. We plot the fitted results in Fig. \ref{fig:lp}(b), we can see that it bears similarities as the mean pore size. 

Moreover, if we interpret Debye random media as overlapping particles of ``random shape and size", we may heuristically relate our results to the one for overlapping polydisperse disks \cite{lu1992lineal2}: 
\begin{equation}
L(z)=\phi_1^{1+2\langle \mathcal{R} \rangle z/(\pi\langle{\mathcal R^2}\rangle)},
\label{eq:poly}
\end{equation}   
where $\langle{\mathcal R}\rangle$ and $\langle {\mathcal R^2} \rangle$ are the first and second moments of the particle size distribution function. The ratio $\langle{\mathcal R^2}\rangle/\langle \mathcal{R} \rangle$ is simply related to $L_w$ by the relation $L_w=-\pi/(2\ln\phi_1)\langle{\mathcal R^2}\rangle/\langle \mathcal{R} \rangle$. Surprisingly, we find that $\langle{\mathcal R^2}\rangle/\langle \mathcal{R} \rangle$ is rather insensitive to the change of volume fractions. Specifically, we find its value is approximately $(0.94\pm0.04)a$, which shows that the lineal-path function of Debye random media is actually quite similar to that of overlapping disk systems with the ratio $\langle{\mathcal R^2}\rangle/\langle \mathcal{R} \rangle$ (reduces to $R$ for monodisperse systems) comparable to the characteristic length scale. 

\subsection{Chord-Length Probability Density Function}
\begin{figure}[h]
\centering
\includegraphics[width=8cm, height=5cm, clip=true]{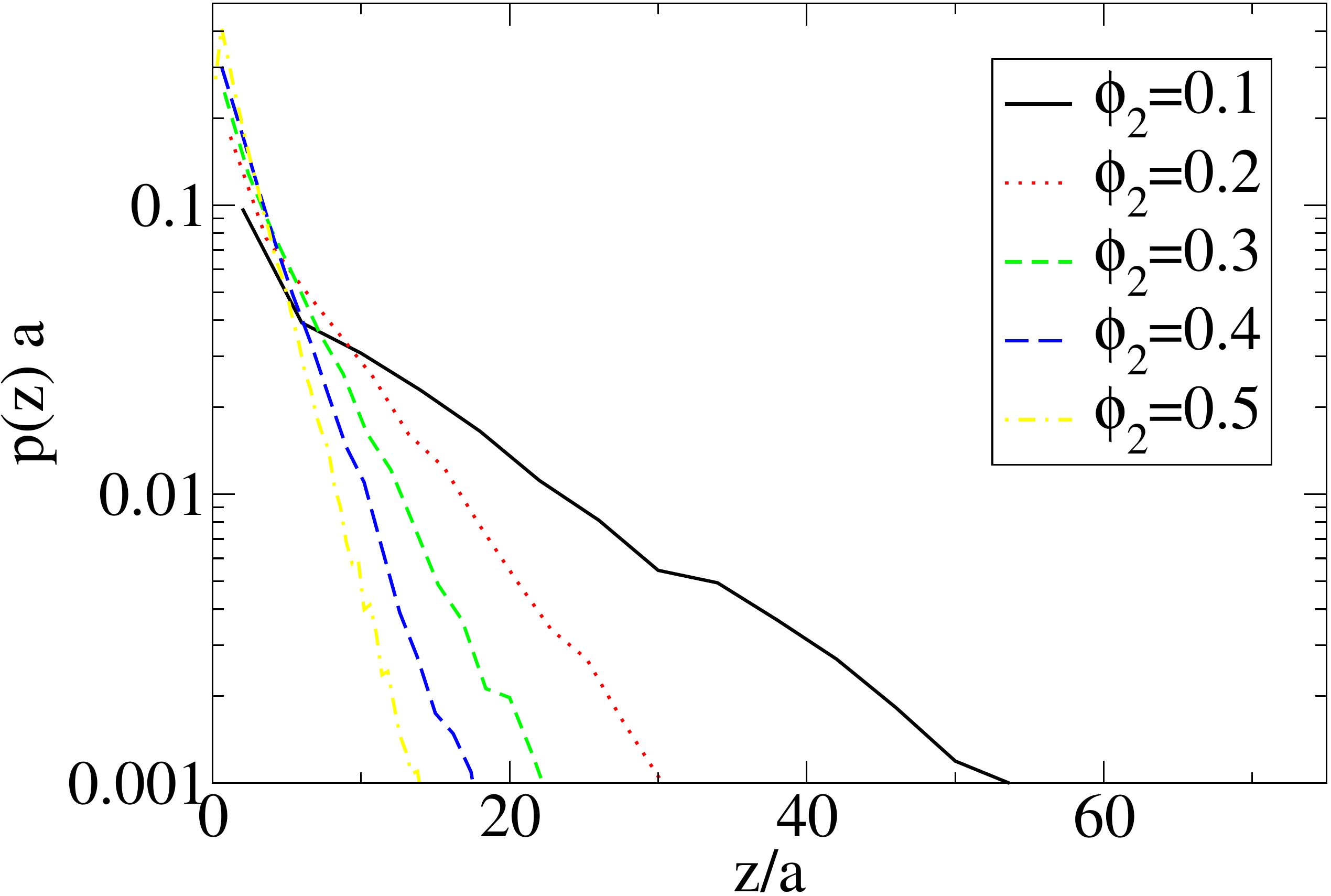}
\caption{The matrix chord-length probability density function $p(z)$ for Debye random media.}
\label{fig: chord}
\end{figure}

Using relations (\ref{eq:pzlz}) and (\ref{eq:chords2}), we can easily obtain the matrix chord-length probability density function $p(z)$ from the second derivative of the lineal-path function $L(z)$ or by direct sampling of the realizations. The simulated results for $p(z)$ for different volume fractions are shown in Fig. \ref{fig: chord}. Clearly, the second derivative of an exponential function is still an exponential function with the same slope on a semi-logarithm plot. Specifically, using Eq. (\ref{eq:lpfit}) we can write the $p(z)$ explicitly:
\begin{equation}
p(z)=\frac{a}{\phi_2L_w^2}\exp(-z/L_w).
\end{equation}
Indeed, we can observe the similarities between Fig. \ref{fig:lp}(a) and Fig. \ref{fig: chord}.  

\section{Comparison to Models of Particle Dispersions}  
\begin{figure}[]
\centering
\subfigure[]{
\includegraphics[clip=true, width=4cm, height=4cm, clip=true]{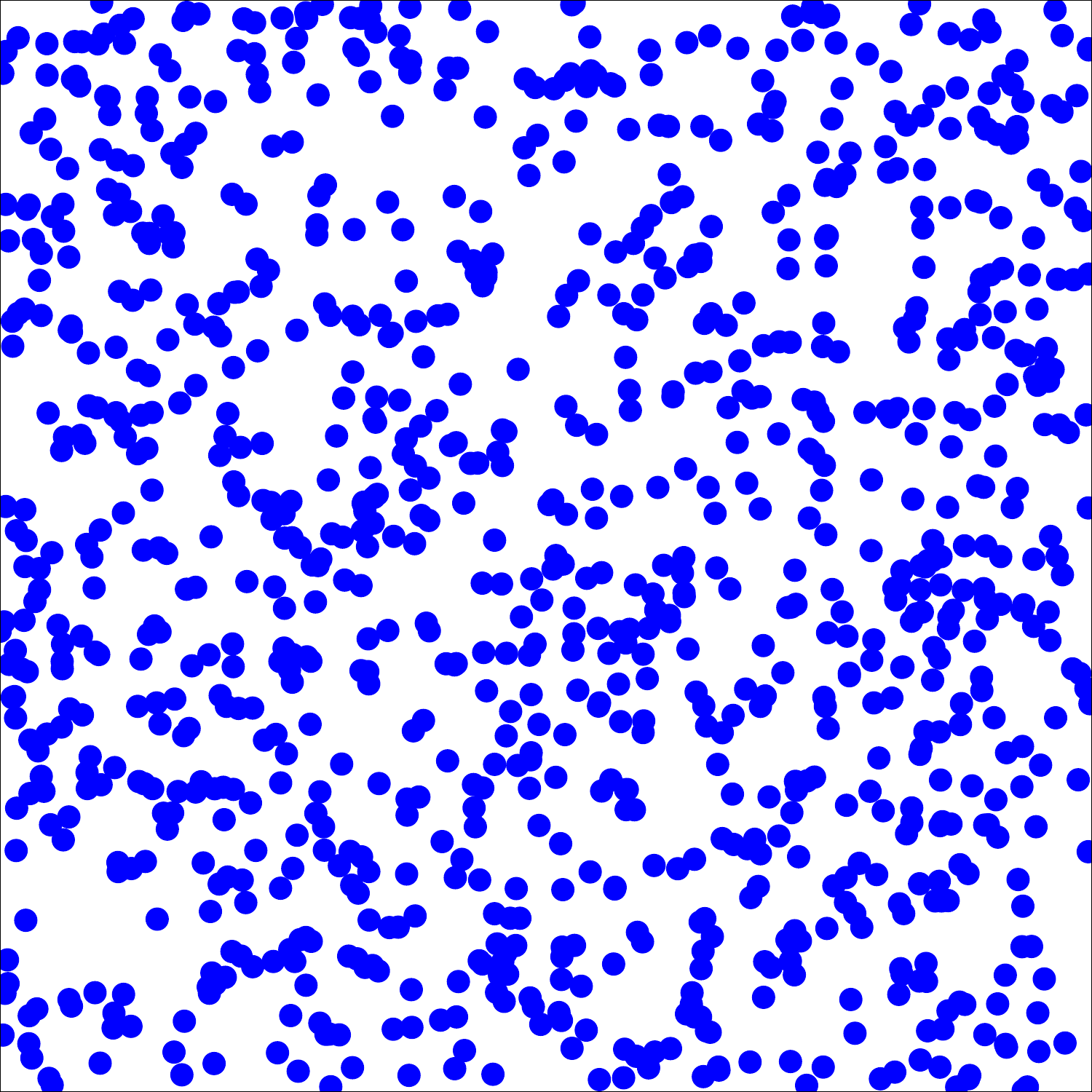}
}
\subfigure[]{
\includegraphics[clip=true, width=4cm, height=4cm, clip=true]{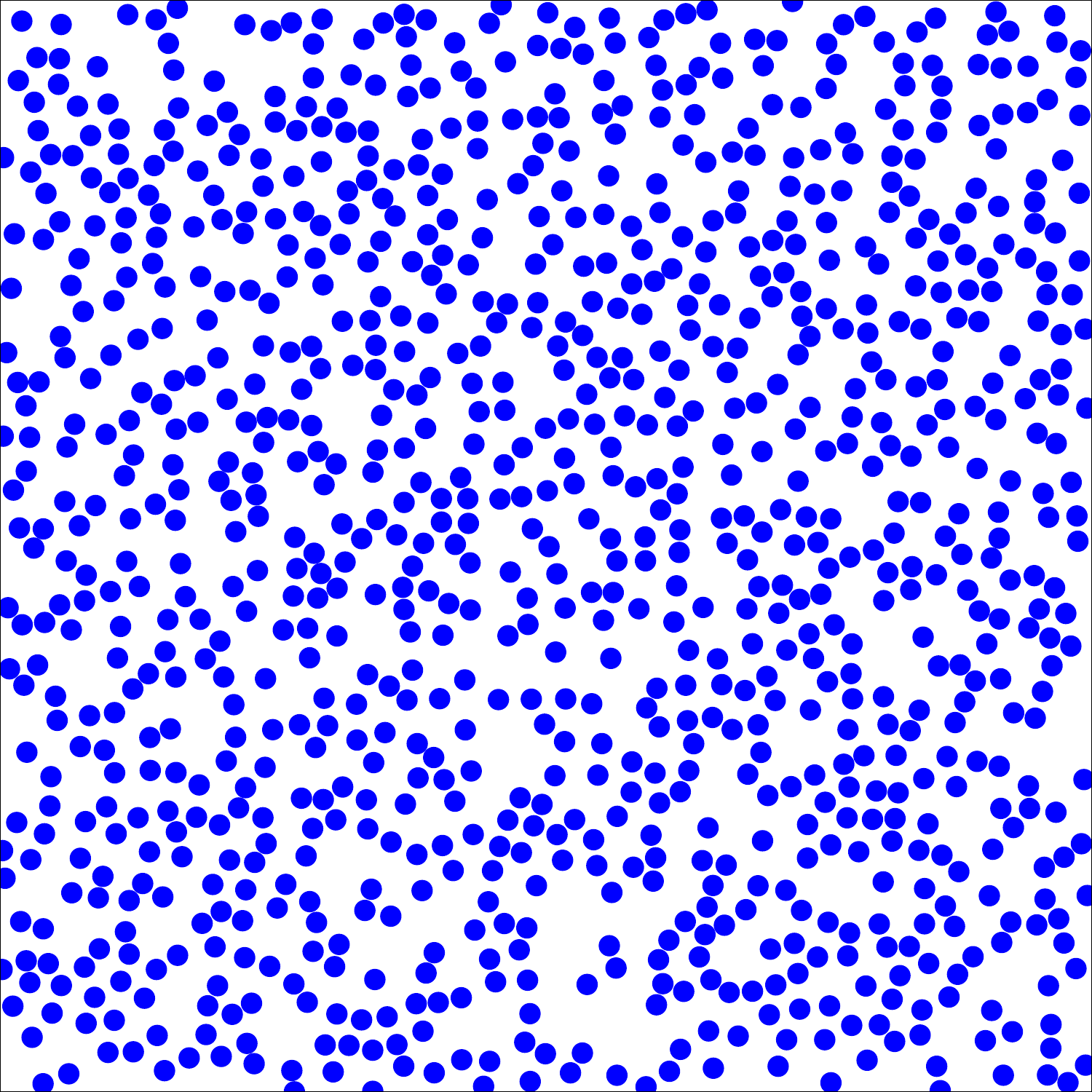}
}
\caption{Representative images of the two models of particle dispersions (each with $\phi_2=0.3$) discussed in Sec. VI. (a) Overlapping disks (b) Equilibrium hard disks.}
\label{fig:particle}
\end{figure}

It is instructive to compare all of the statistical descriptors
considered here for Debye random
media to corresponding results for
models of particle dispersions. Specifically, we consider overlapping
disks as well as equilibrium hard disks. Neither of these models have phase-inversion symmetry. Representative images of both systems are shown in Fig. (\ref{fig:particle}).

\subsection{Overlapping Disks}
Overlapping spheres (also called fully-penetrable sphere model) refer to an uncorrelated (Poisson) distribution of spheres of radius $R$ throughout a matrix \cite{torquato2013random, To86i}. It is also a special case of the ``Boolean model" known in stochastic geometry \cite{St95}. In principle, the $n$-point correlation function for this model can be determined exactly \cite{torquato2013random}. In $d$-dimensional Euclidean space $\mathbb{R}^d$, the volume fraction of the void phase $\phi_1$ is given by $\exp(-\rho v_1(R))$, where $\rho$ is the number density and $v_1(R)$ is the volume of a sphere of radius $R$. The two-point correlation function $S_2^{(1)}(r)$ is given by
\begin{equation}
S_2^{(1)}(r)=\exp(-\rho v_2(r;R)),
\label{eq:overlappings2}
\end{equation} 
where $v_2(r;R)$ represents the union volume of two spheres whose centers are separated by a distance $r$. In this paper we are particularly interested in $d=2$, in which case $v_2(r;R)$ can be explicitly written as 
\begin{equation}
\begin{split}
&\frac{v_2(r;R)}{v_1(R)}=2\Theta(r-2R)\\&+\frac{2}{\pi}\left[ \pi+\frac{r}{2R}(1-\frac{r^2}{4R^2})^{1/2}-\cos^{-1}(\frac{r}{2R})\right]\Theta(2R-r),
\end{split}
\end{equation} 
where $\Theta(x)$ is the Heaviside step function and $v_1(R)=\pi R^2$. 

The specific surface $s$ for overlapping disks is simply given by 
\begin{equation}
s=\frac{2\eta\phi_1}{R}, 
\label{eq:overlaps}
\end{equation}
where $\eta=\rho v_1(R)=-\ln(\phi_1)$. We also derive explicit expressions for two-point surface correlation functions for overlapping disks following the procedures detailed in Ref. \cite{torquato2013random}, which to our knowledge have not been reported elsewhere. Specifically, the surface-void correlation function is given by
\begin{equation}
F_{\mathrm{sv}}(r)=\frac {2\eta}{R}[1-\frac{1}{\pi}\cos^{-1}(\frac{r}{2R})\Theta(2R-r)]S_2^{(1)}(r),
\label{eq:2dfsv}
\end{equation}  
where $S_2^{(1)}(r)$ is already given in Eq. (\ref{eq:overlappings2}). The surface-surface correlation function is given by
\begin{equation}
\begin{split}
F_{\mathrm{ss}}(r)=S_2^{(1)}(r)\{\frac {4\eta^2}{R^2}[1-\frac{1}{\pi}\cos^{-1}(\frac{r}{2R})\Theta(2R-r)]^2\\
+\frac{2\eta}{\pi R}\frac{1}{r(1-(\frac{r}{2R})^2)^{1/2}}\Theta(2R-r)\},
\end{split}
\label{eq:2dfss}
\end{equation}  
note that $F_{\mathrm{ss}}(r)$ diverges at $r=2R$.

Other microstructural descriptors mentioned in Sec. II can also be obtained analytically for overlapping disks \cite{torquato2013random}. The pore-size probability density function $P(\delta)$ is given by
\begin{equation}
P(\delta)=\frac{2\eta}{R}(1+\frac{\delta}{R})\phi_1^{\delta^2/R^2+2\delta/R}.
\label{eq:overlappore}
\end{equation}
The lineal-path function $L(z)$ for the matrix phase is simply given by
\begin{equation}
L(z)=\phi_1^{\textstyle 1+\frac{2z}{\pi R}}.
\label{eq:overlaplz}
\end{equation}
The matrix chord-length density function $p(z)$ is given by
\begin{equation}
p(z)=\frac{2\eta}{\pi R}\phi_1^{\textstyle\frac{2z}{\pi R}}.
\label{eq:overlapchord}
\end{equation}

\subsection{Equilibrium Hard Disks}
We also consider distributions of identical hard disks of radius $R$ in equilibrium (Gibbs ensemble) along the stable fluid branch \cite{torquato2013random,hansen1990theory}. The correlation functions of this model are directly related to integrals over their pair correlation functions \cite{torquato1985microstructure,To86i}, which
can be estimated via the Percus-Yevick approximation \cite{torquato2013random}, which is however only analytically solvable for odd dimensions. The specific surface is simply given by
\begin{equation}
s=\frac{2\phi_2}{R}. 
\label{eq:hards}
\end{equation}
Here we obtain its two-point correlation functions [$\chi_{_V}(r)$, $F_{sv}(r)$ and $F_{ss}(r)$] from Monte Carlo simulations. 
For the pore-size function, lineal-path function and the chord-length density function for equilibrium hard disks, we use the excellent analytical approximations \cite{torquato2013random,torquato1995nearest,lu1992lineal}. The expression for $P(\delta)$ is given by 
\begin{equation}
P(\delta)=\frac{4\phi_2}{R}(a_0x+a_1)\exp\left [-\phi_2(4a_0x^2+8a_1x+a_2)\right ],
\label{eq:hardpore}
\end{equation}
where $x=\delta/(2R)+1/2$ and $a_0, a_1, a_2$ are volume-fraction dependent coefficients given in Ref. \cite{torquato2013random}. The lineal-path function $L(z)$ for the matrix phase is given by \cite{lu1992lineal} 
\begin{equation}
L(z)=\phi_1\exp(-\frac{2\phi_2}{\pi\phi_1}\frac{z}{R}).
\label{eq:hardlp}
\end{equation}
The matrix chord-length density function $p(z)$ is then given by   
\begin{equation}
p(z)=\frac{2\phi_2}{\pi\phi_1R}\exp(-\frac{2\phi_2}{\pi\phi_1}\frac{z}{R}).
\label{eq:hardchord}
\end{equation}

\begin{figure}[]
\centering
\subfigure[]{
\includegraphics[width=7cm, height=5cm, clip=true]{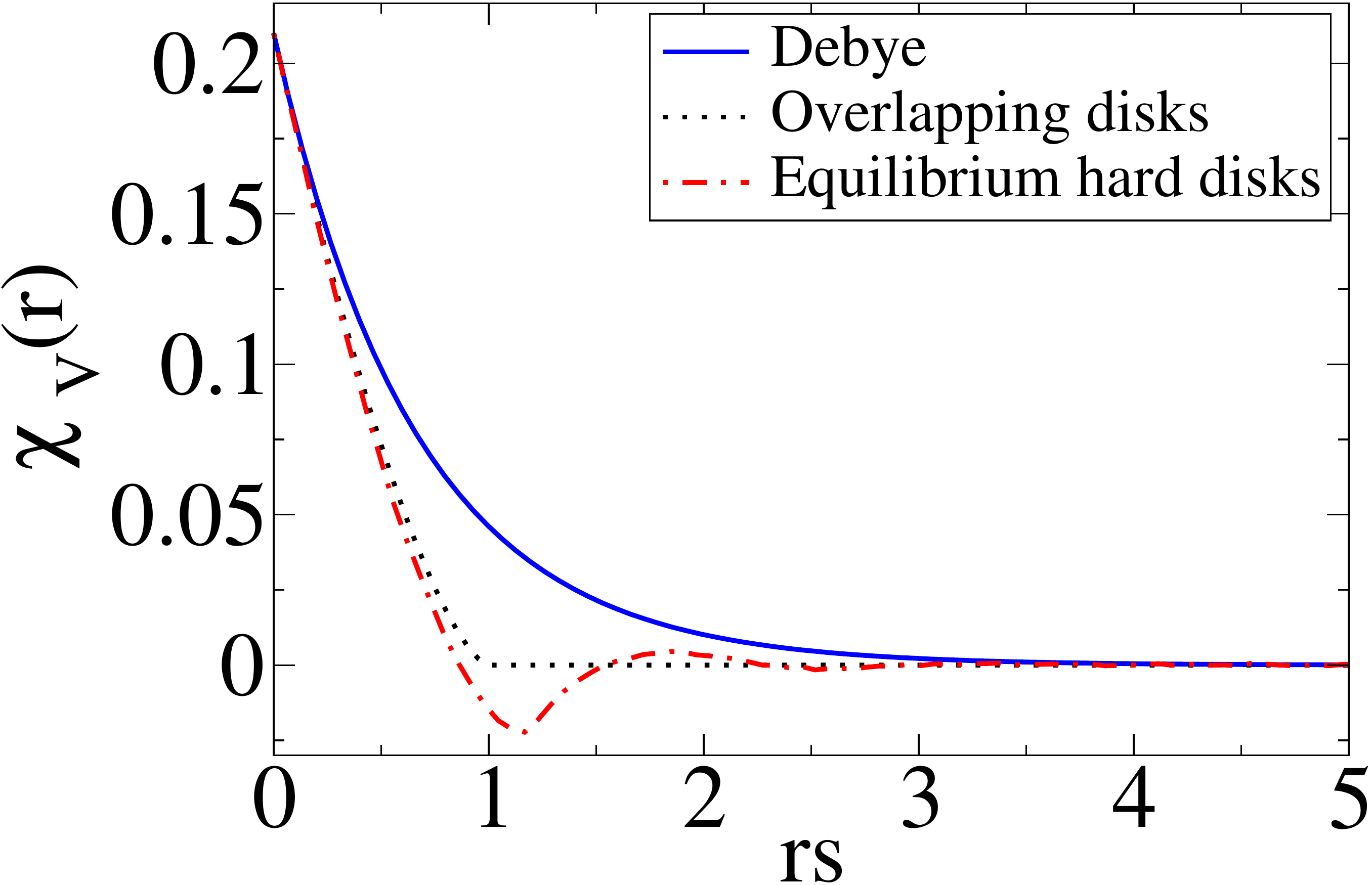}
}
\subfigure[]{
\includegraphics[width=7cm, height=5cm, clip=true]{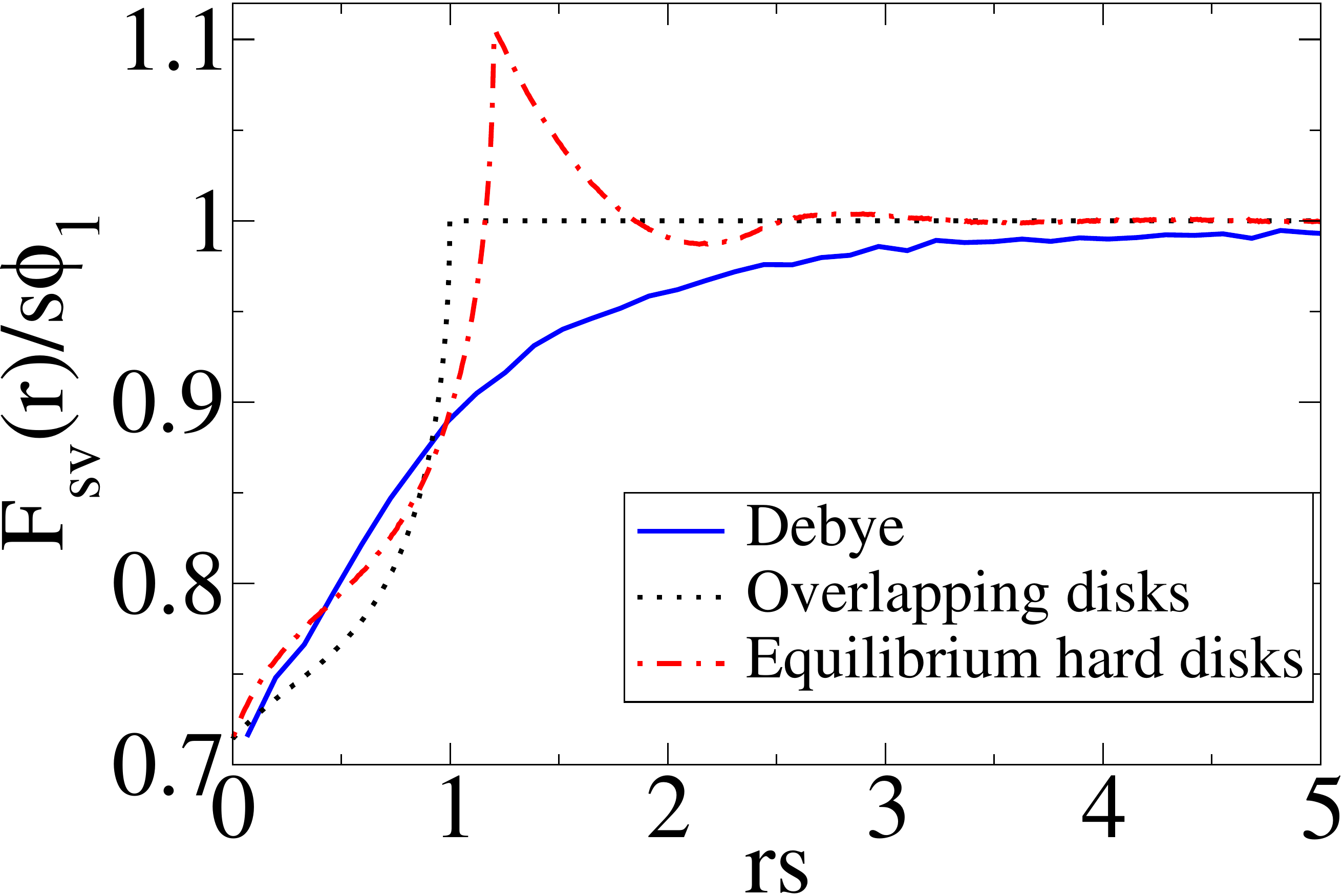}
}
\subfigure[]{
\includegraphics[width=7cm, height=5cm, clip=true]{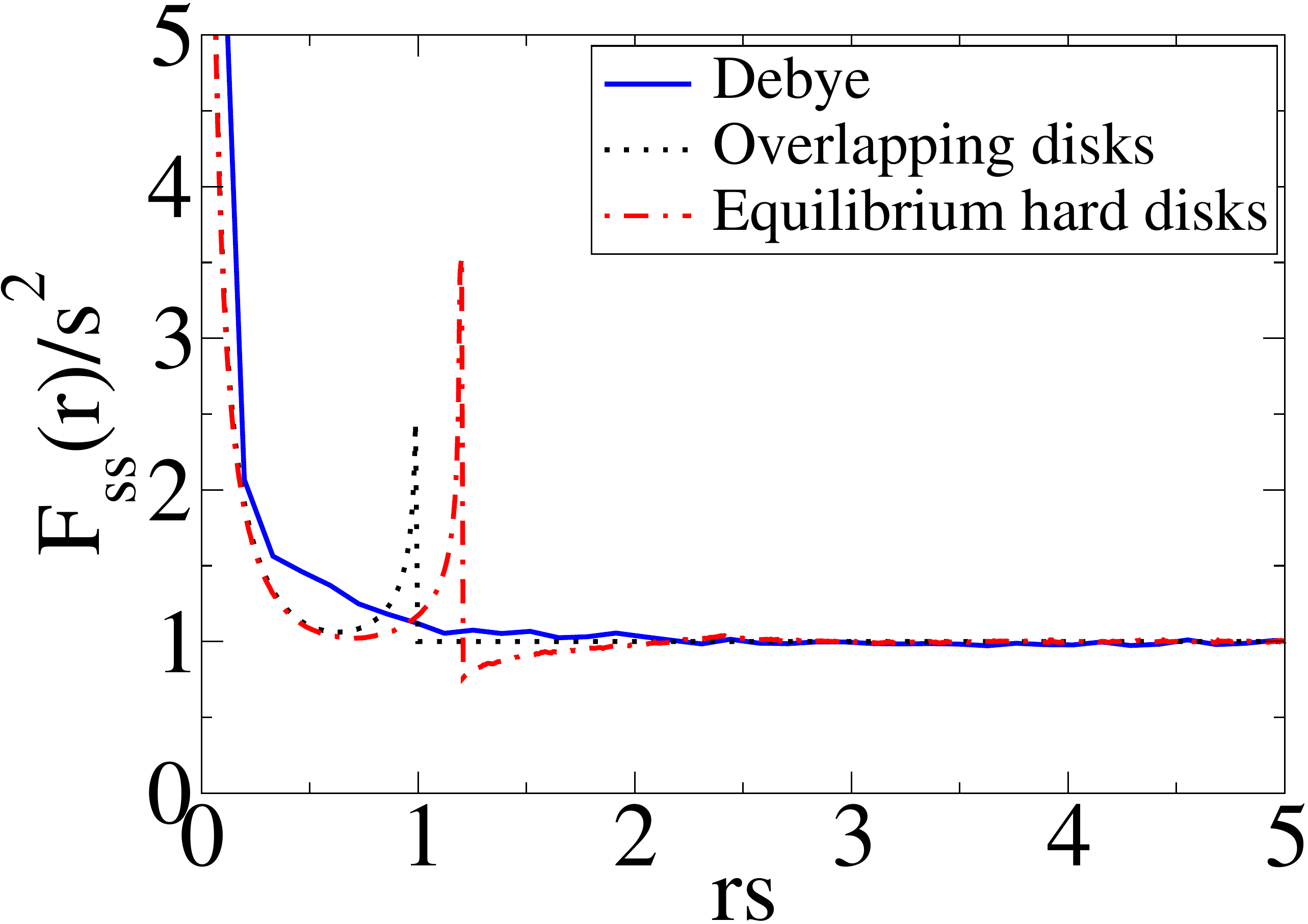}
}
\caption{Comparison of (a) autocovariance function $\chi_{_{V}}(r)$, (b) surface-void correlation function $F_{\mathrm{sv}}(r)$ and (c) surface-surface correlation function $F_{\mathrm{ss}}(r)$ for Debye random media, overlapping disks and equilibrium hard disks at $\phi_2=0.3$. The radial distance $r$ is scaled by the specific surface $s$.}
\label{fig:two-point-compare}
\end{figure}

\subsection{Results} 
We evaluate and compare the aforementioned statistical descriptors for Debye random media, overlapping disks and equilibrium hard disks at a fixed volume fraction (here we use $\phi_2=0.3$ as it lies in the middle of the volume fractions we target for constructions). Analytical expressions are used whenever they are available.\\
\indent In Fig. \ref{fig:two-point-compare} we compare the autocovariance function $\chi_{_{V}}(r)$, surface-void correlation function $F_{\mathrm{sv}}(r)$ and surface-surface correlation function $F_{\mathrm{ss}}(r)$ for three models. The surface correlation functions are scaled by their large-$r$ values for the convenience of comparison. As the characteristic length scale $a$ and particle radius $R$ are only defined for their corresponding models, we use the specific surface $s$ to scale the distance $r$. The scaled distance can be related back to $a$ and $R$ via Eq. (\ref{eq:s}), Eq. (\ref{eq:overlaps}) and Eq. (\ref{eq:hards}) for Debye random media, overlapping disks and equilibrium hard disks, respectively. It can be seen that overlapping disks are uncorrelated when $r \ge 2R$. By contrast, equilibrium hard disks exhibit positive and negative correlations and remain correlated beyond $2R$. Debye random media turns to have the most persistent correlations within the length scale shown in Fig. \ref{fig:two-point-compare}. Interestingly, Debye random media exhibit monotonic behaviors for all three correlation functions. In the case of autocovariance function, this means only positive correlations. Most importantly, the correlation functions of Debye random media are all smooth and free from non-differentiable kinks. The absence of these discontinuities, which are marked by the particle diameter as in the cases of the other two disk systems, implies an absence of regular domains in Debye random media.
\begin{figure}[h]
\centering
\includegraphics[width=8cm, height=5cm, clip=true]{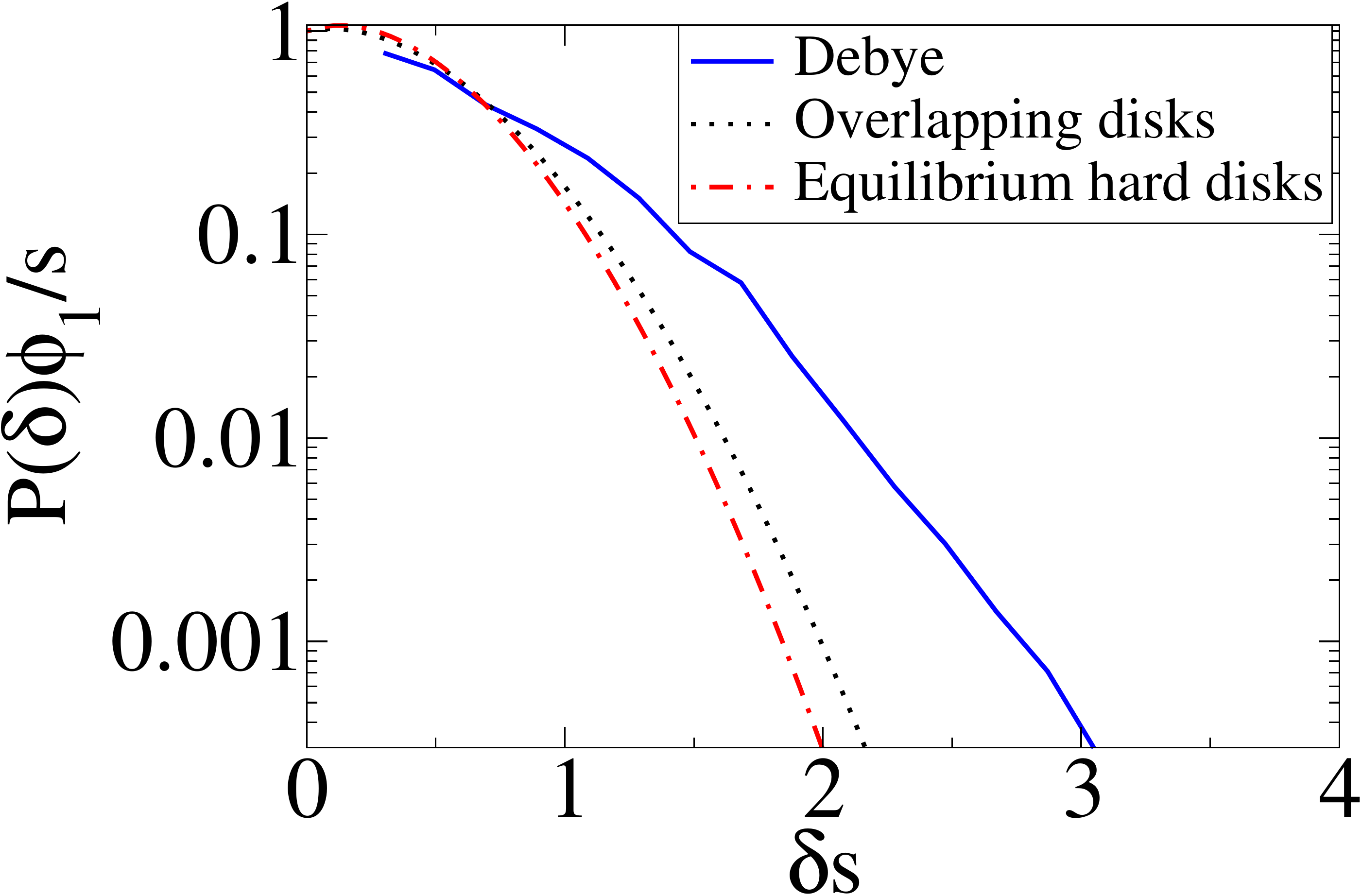}
\caption{The pore-size probability density function $P(\delta)$ for three different systems, Debye random media, overlapping disks and equilibrium hard disks at $\phi_2=0.3$. The pore-size distance $\delta$ is scaled by the specific surface $s$.}
\label{fig:pore_compare}
\end{figure}

\begin{figure}[h]
\centering
\includegraphics[width=8cm, height=5cm, clip=true]{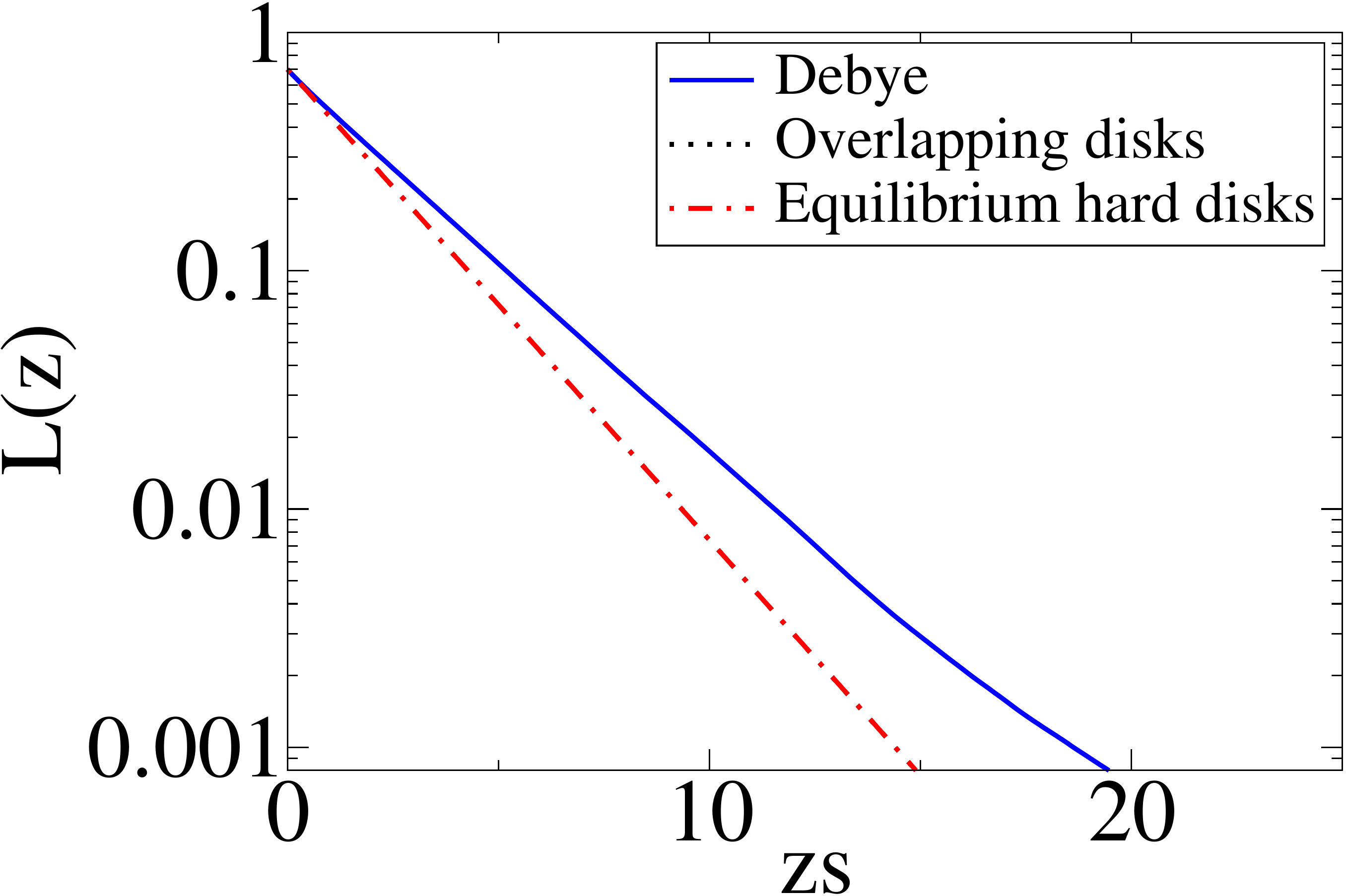}
\caption{The lineal-path function $L(z)$ for three different systems, Debye random media, overlapping disks and equilibrium hard disks at $\phi_2=0.3$. Note that 
curves for $L(z)$ for overlapping disks and equilibrium hard disks are the same after we scale $z$ by $s$.}
\label{fig:lp_compare}
\end{figure}
   
\begin{figure}[h]
\centering
\includegraphics[width=8cm, height=5cm, clip=true]{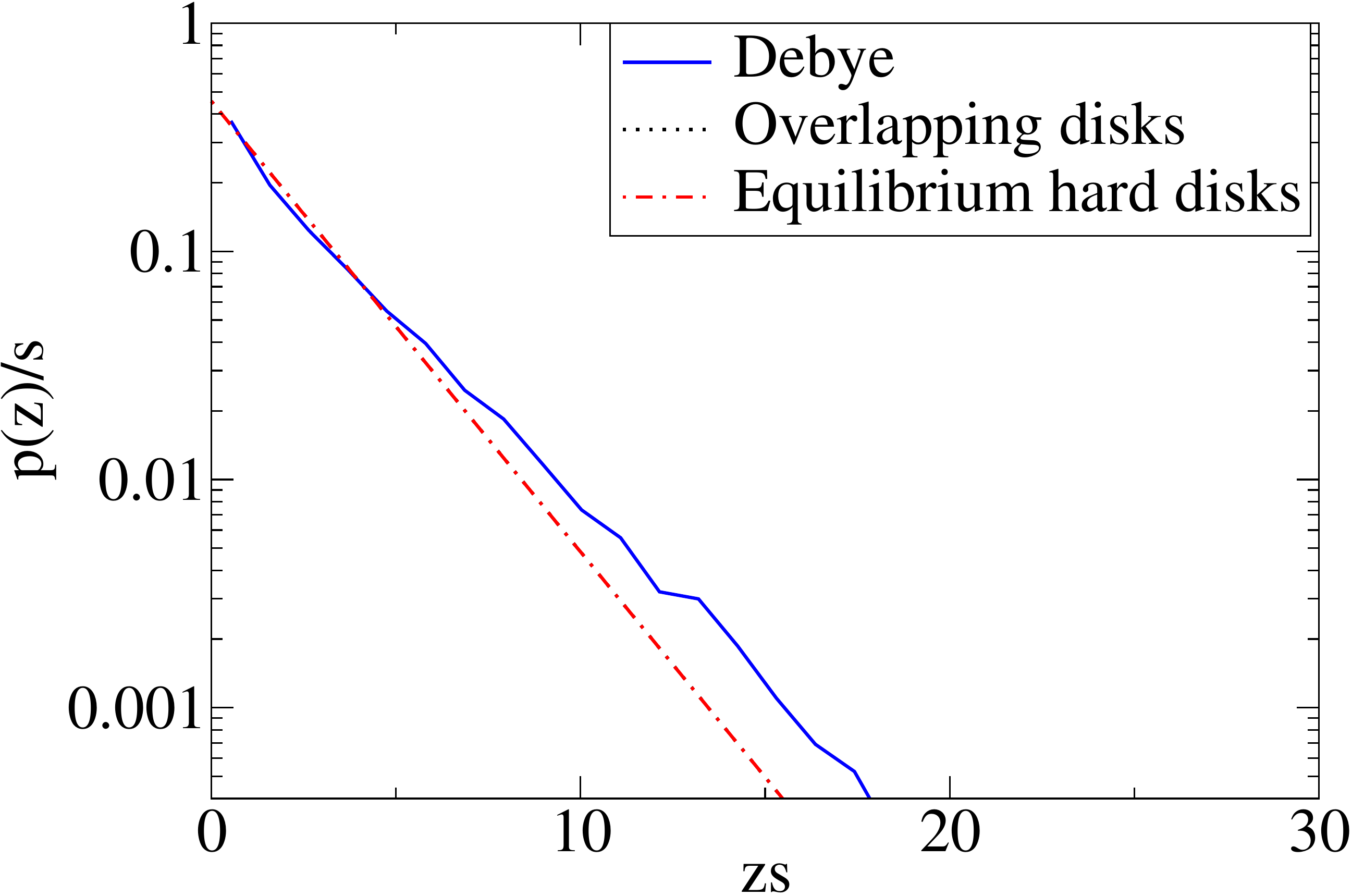}
\caption{The matrix chord-length density function $p(z)$ for three different systems, Debye random media, overlapping disks and equilibrium hard disks at $\phi_2=0.3$.
Note that by scaling $z$ by $s$, the curves for $p(z)$ for overlapping disks and equilibrium hard disks are the same.}
\label{fig:pz_compare}
\end{figure}

In Fig. \ref{fig:pore_compare}, we compare the pore-size probability density function $P(\delta)$ for three different systems, Debye random media, overlapping disks and equilibrium hard disks at volume fraction $\phi_2=0.3$. The tail of $P(\delta)$ is a measure of how likely it is to find a large hole. We find that $P(\delta)$ of the overlapping disks is larger than that of equilibrium hard disks. This can be understood by noting the fact that particles are more likely to cluster in overlapping disks, which leads to larger void domains. Interestingly, we observe that $P(\delta)$ of Debye random media is much larger than both disk models for large $\delta$. This indeed confirms that Debye random media possesses a significant fraction of large holes, as suggested in Ref. \cite{torquato2020predicting}. This is because the autocovariance function of Debye random media has infinite support, whereas the one for overlapping spheres has finite support; see Ref. \cite{torquato2020predicting}.\\
\indent We also compare the lineal-path function $L(z)$ for these three systems, as shown in Fig. \ref{fig:lp_compare}. We again find that the lineal-path function of Debye random media decays slower than those of the other two. This is consistent with ``large-hole" property obtained
from pore-size probability density function $P(\delta)$, since $L(z)$ measures the probability of an entire line of length $z$ lying in the pore phase. However, we notice that the difference between the lineal-path functions is much less prominent than that of pore-size probability density function. Interestingly, the lineal-path functions for overlapping disks (see Eq. (\ref{eq:overlaplz})) and equilibrium hard disks \cite{lu1992lineal} (see Eq. (\ref{eq:hardlp})) are the same after we scale $z$ by the specific surface $s$. This suggests that the pore-size probability density function is a more sensitive measure of the pore space compared to the lineal-path function. \\
\indent Finally, we compare the matrix chord-length density function $p(z)$ for three systems, as shown in Fig. \ref{fig:pz_compare}. The results for overlapping disks and equilibrium hard disks are obtained via Eq. (\ref{eq:overlapchord}) and Eq. (\ref{eq:hardchord}). We see that $p(z)$ for Debye random media decays slower, implying that larger chords have larger weights than those for the other models, which is consistent with the argument concerning the existence of large ``holes" \cite{torquato2020predicting}. 
\section{Conclusions and Discussion}  
In this work, we have constructed a class of 2D Debye random media using an accelerated Yeong-Torquato construction algorithm and study its microstructural descriptors. Specifically, we compute the two-point correlation functions, pore-size functions, lineal-path function, and chord-length probability density function. Importantly, we devised accurate semi-analytical and empirical formulas for these descriptors. By comparing these results to those of overlapping disks and equilibrium hard disks, we find that all three two-point correlation functions for Debye random media are monotonic with the distance $r$ and are more long-ranged than those of the particle dispersion models. The absence of discontinuities in the two-point correlation functions means that there is no unique domain size for Debye random media, which is consistent with Debye's intuition that these domains consist of ``random shapes and sizes". On the other hand, results for the pore-size functions, lineal-path function, and chord-length probability density function show that Debye random media possess large ``holes" compared to overlapping disks and equilibrium hard disks, as suggested in Ref. \cite{torquato2020predicting}.

Our detailed structural characterization of Debye random media for $d=2$ has implications on its properties for $d=3$. Importantly, the demonstration that 2D Debye random media tend to possess a wide spectrum of hole sizes, including a substantial fraction of large holes is expected to be true for 3D random media. Indeed, a small sample of a 3D Debye random media reported in Ref. \cite{torquato2020predicting} bears this out. However, the construction of 3D Debye random media of large sizes corresponding to the 2D one reported here ($500^3$ voxels), even with the fast algorithm presented here, is still a challenging computational task ($\sim10^5$ computing hours for an Intel Core i5 processor). Thus, the development of efficient algorithms for constructing 3D Debye random media is an outstanding problem for future research. Specifically, we expect that our proposed semi-analytical expressions for surface correlation functions will hold in higher dimensions, given their very general forms. 

In three dimensions, by replacing the specific surface with the corresponding expression in Eq. (\ref{eq:debyeFsv}) and Eq. (\ref{eq:debyeFss}), we make the following proposals for approximation formulas for the surface correlation functions for 3D Debye random media:
\begin{equation}
F_{\mathrm{sv}}(r)=\frac{4\phi_2}{a}\frac{1}{1+\exp(-r/a)}S_2^{(1)}(r),
\end{equation} 
and
\begin{equation} 
\begin{split}
F_{\mathrm{ss}}(r)=\frac{16}{a^2}\phi_1^2\phi_2^2+\frac{2}{ar}\phi_1\phi_2\exp(-r/a)\\+\frac{1}{2a^2}\frac{\exp(-r/a)}{1+\exp(-r/a)}|\phi_2-\phi_1|.
\end{split}
\end{equation}
It is also reasonable to deduce that the lineal-path function and chord-length probability density function for 3D Debye random media will have exponential forms that resemble those of overlapping spheres.  

Moreover, it would be of great interest to estimate the degeneracy of 
Debye random media with the autocovariance (\ref{S2Debye}) using the techniques in Ref. \cite{gommes2012density}. Specifically, it would be desirable to specifically generate Debye random media that lie outside the ``most probable" class studied here. This could be done by biasing the construction algorithm to have an energy that targets not only (\ref{S2Debye}) but also other
microstructural descriptors, as was done in Ref. \cite{yeong1998reconstructing}.  

Finally, we note that our fast implementation
of the Yeong-Torquato algorithm can be applied to study other disordered
microstructures whose autocovariance function decays sufficiently fast. In particular, it can be used to generate other models
defined by their two-point correlation functions \cite{jiao2008modeling} and study other microstructural descriptors
of such media.

\begin{acknowledgements}
\indent The authors gratefully acknowledge the support of the National 
Science Foundation under Grant No. CBET-1701843.
\end{acknowledgements}


\begin{thebibliography}{42}%
\makeatletter
\providecommand \@ifxundefined [1]{%
 \@ifx{#1\undefined}
}%
\providecommand \@ifnum [1]{%
 \ifnum #1\expandafter \@firstoftwo
 \else \expandafter \@secondoftwo
 \fi
}%
\providecommand \@ifx [1]{%
 \ifx #1\expandafter \@firstoftwo
 \else \expandafter \@secondoftwo
 \fi
}%
\providecommand \natexlab [1]{#1}%
\providecommand \enquote  [1]{``#1''}%
\providecommand \bibnamefont  [1]{#1}%
\providecommand \bibfnamefont [1]{#1}%
\providecommand \citenamefont [1]{#1}%
\providecommand \href@noop [0]{\@secondoftwo}%
\providecommand \href [0]{\begingroup \@sanitize@url \@href}%
\providecommand \@href[1]{\@@startlink{#1}\@@href}%
\providecommand \@@href[1]{\endgroup#1\@@endlink}%
\providecommand \@sanitize@url [0]{\catcode `\\12\catcode `\$12\catcode
  `\&12\catcode `\#12\catcode `\^12\catcode `\_12\catcode `\%12\relax}%
\providecommand \@@startlink[1]{}%
\providecommand \@@endlink[0]{}%
\providecommand \url  [0]{\begingroup\@sanitize@url \@url }%
\providecommand \@url [1]{\endgroup\@href {#1}{\urlprefix }}%
\providecommand \urlprefix  [0]{URL }%
\providecommand \Eprint [0]{\href }%
\providecommand \doibase [0]{http://dx.doi.org/}%
\providecommand \selectlanguage [0]{\@gobble}%
\providecommand \bibinfo  [0]{\@secondoftwo}%
\providecommand \bibfield  [0]{\@secondoftwo}%
\providecommand \translation [1]{[#1]}%
\providecommand \BibitemOpen [0]{}%
\providecommand \bibitemStop [0]{}%
\providecommand \bibitemNoStop [0]{.\EOS\space}%
\providecommand \EOS [0]{\spacefactor3000\relax}%
\providecommand \BibitemShut  [1]{\csname bibitem#1\endcsname}%
\let\auto@bib@innerbib\@empty
\bibitem [{\citenamefont {Torquato}(2002)}]{torquato2013random}%
  \BibitemOpen
  \bibfield  {author} {\bibinfo {author} {\bibfnamefont {S.}~\bibnamefont
  {Torquato}},\ }\href@noop {} {\emph {\bibinfo {title} {Random Heterogeneous
  Materials: Microstructure and Macroscopic Properties}}}\ (\bibinfo
  {publisher} {Springer Science \& Business Media},\ \bibinfo {year}
  {2002})\BibitemShut {NoStop}%
\bibitem [{\citenamefont {Milton}(2002)}]{Mi02}%
  \BibitemOpen
  \bibfield  {author} {\bibinfo {author} {\bibfnamefont {G.~W.}\ \bibnamefont
  {Milton}},\ }\href@noop {} {\emph {\bibinfo {title} {The Theory of
  Composites}}}\ (\bibinfo  {publisher} {Cambridge University Press},\ \bibinfo
  {address} {Cambridge, England},\ \bibinfo {year} {2002})\BibitemShut
  {NoStop}%
\bibitem [{\citenamefont {Sahimi}(2003)}]{sahimi2003heterogeneous}%
  \BibitemOpen
  \bibfield  {author} {\bibinfo {author} {\bibfnamefont {M.}~\bibnamefont
  {Sahimi}},\ }\href@noop {} {\emph {\bibinfo {title} {Heterogeneous Materials
  I: Linear transport and optical properties}}},\ Vol.~\bibinfo {volume} {22}\
  (\bibinfo  {publisher} {Springer Science \& Business Media},\ \bibinfo {year}
  {2003})\BibitemShut {NoStop}%
\bibitem [{\citenamefont {Patel}\ and\ \citenamefont {Zohdi}(2016)}]{Pa16}%
  \BibitemOpen
  \bibfield  {author} {\bibinfo {author} {\bibfnamefont {B.}~\bibnamefont
  {Patel}}\ and\ \bibinfo {author} {\bibfnamefont {T.~I.}\ \bibnamefont
  {Zohdi}},\ }\href@noop {} {\bibfield  {journal} {\bibinfo  {journal} {Mater.
  Des.}\ }\textbf {\bibinfo {volume} {94}},\ \bibinfo {pages} {546} (\bibinfo
  {year} {2016})}\BibitemShut {NoStop}%
\bibitem [{\citenamefont {Hristopulos}(2020)}]{hristopulos2020random}%
  \BibitemOpen
  \bibfield  {author} {\bibinfo {author} {\bibfnamefont {D.~T.}\ \bibnamefont
  {Hristopulos}},\ }\href@noop {} {\emph {\bibinfo {title} {Random Fields for
  Spatial Data Modeling}}}\ (\bibinfo  {publisher} {Springer},\ \bibinfo {year}
  {2020})\BibitemShut {NoStop}%
\bibitem [{\citenamefont {Gibson}\ and\ \citenamefont
  {Ashby}(1999)}]{gibson1999cellular}%
  \BibitemOpen
  \bibfield  {author} {\bibinfo {author} {\bibfnamefont {L.~J.}\ \bibnamefont
  {Gibson}}\ and\ \bibinfo {author} {\bibfnamefont {M.~F.}\ \bibnamefont
  {Ashby}},\ }\href@noop {} {\emph {\bibinfo {title} {Cellular solids:
  structure and properties}}}\ (\bibinfo  {publisher} {Cambridge University
  Press},\ \bibinfo {year} {1999})\BibitemShut {NoStop}%
\bibitem [{\citenamefont {Wadsworth}\ \emph {et~al.}(2016)\citenamefont
  {Wadsworth}, \citenamefont {Vasseur}, \citenamefont {Scheu}, \citenamefont
  {Kendrick}, \citenamefont {Lavall{\'e}e},\ and\ \citenamefont
  {Dingwell}}]{wadsworth2016universal}%
  \BibitemOpen
  \bibfield  {author} {\bibinfo {author} {\bibfnamefont {F.~B.}\ \bibnamefont
  {Wadsworth}}, \bibinfo {author} {\bibfnamefont {J.}~\bibnamefont {Vasseur}},
  \bibinfo {author} {\bibfnamefont {B.}~\bibnamefont {Scheu}}, \bibinfo
  {author} {\bibfnamefont {J.~E.}\ \bibnamefont {Kendrick}}, \bibinfo {author}
  {\bibfnamefont {Y.}~\bibnamefont {Lavall{\'e}e}}, \ and\ \bibinfo {author}
  {\bibfnamefont {D.~B.}\ \bibnamefont {Dingwell}},\ }\href@noop {} {\bibfield
  {journal} {\bibinfo  {journal} {Geology}\ }\textbf {\bibinfo {volume} {44}},\
  \bibinfo {pages} {219} (\bibinfo {year} {2016})}\BibitemShut {NoStop}%
\bibitem [{\citenamefont {Torquato}(1986)}]{To86i}%
  \BibitemOpen
  \bibfield  {author} {\bibinfo {author} {\bibfnamefont {S.}~\bibnamefont
  {Torquato}},\ }\href@noop {} {\bibfield  {journal} {\bibinfo  {journal} {J.
  Stat. Phys.}\ }\textbf {\bibinfo {volume} {45}},\ \bibinfo {pages} {843}
  (\bibinfo {year} {1986})}\BibitemShut {NoStop}%
\bibitem [{\citenamefont {Stoyan}\ \emph {et~al.}(1995)\citenamefont {Stoyan},
  \citenamefont {Kendall},\ and\ \citenamefont {Mecke}}]{St95}%
  \BibitemOpen
  \bibfield  {author} {\bibinfo {author} {\bibfnamefont {D.}~\bibnamefont
  {Stoyan}}, \bibinfo {author} {\bibfnamefont {W.~S.}\ \bibnamefont {Kendall}},
  \ and\ \bibinfo {author} {\bibfnamefont {J.}~\bibnamefont {Mecke}},\
  }\href@noop {} {\emph {\bibinfo {title} {Stochastic Geometry and Its
  Applications}}},\ \bibinfo {edition} {2nd}\ ed.\ (\bibinfo  {publisher}
  {Wiley},\ \bibinfo {address} {New York},\ \bibinfo {year} {1995})\BibitemShut
  {NoStop}%
\bibitem [{\citenamefont {Yeong}\ and\ \citenamefont
  {Torquato}(1998)}]{yeong1998reconstructing}%
  \BibitemOpen
  \bibfield  {author} {\bibinfo {author} {\bibfnamefont {C.~L.~Y.}\
  \bibnamefont {Yeong}}\ and\ \bibinfo {author} {\bibfnamefont
  {S.}~\bibnamefont {Torquato}},\ }\href@noop {} {\bibfield  {journal}
  {\bibinfo  {journal} {Phys. Rev. E}\ }\textbf {\bibinfo {volume} {57}},\
  \bibinfo {pages} {495} (\bibinfo {year} {1998})}\BibitemShut {NoStop}%
\bibitem [{\citenamefont {Debye}\ \emph {et~al.}(1957)\citenamefont {Debye},
  \citenamefont {Anderson~Jr},\ and\ \citenamefont
  {Brumberger}}]{debye1957scattering}%
  \BibitemOpen
  \bibfield  {author} {\bibinfo {author} {\bibfnamefont {P.}~\bibnamefont
  {Debye}}, \bibinfo {author} {\bibfnamefont {H.}~\bibnamefont {Anderson~Jr}},
  \ and\ \bibinfo {author} {\bibfnamefont {H.}~\bibnamefont {Brumberger}},\
  }\href@noop {} {\bibfield  {journal} {\bibinfo  {journal} {J. Appl. Phys.}\
  }\textbf {\bibinfo {volume} {28}},\ \bibinfo {pages} {679} (\bibinfo {year}
  {1957})}\BibitemShut {NoStop}%
\bibitem [{\citenamefont {Coker}\ \emph {et~al.}(1996)\citenamefont {Coker},
  \citenamefont {Torquato},\ and\ \citenamefont
  {Dunsmuir}}]{coker1996morphology}%
  \BibitemOpen
  \bibfield  {author} {\bibinfo {author} {\bibfnamefont {D.~A.}\ \bibnamefont
  {Coker}}, \bibinfo {author} {\bibfnamefont {S.}~\bibnamefont {Torquato}}, \
  and\ \bibinfo {author} {\bibfnamefont {J.~H.}\ \bibnamefont {Dunsmuir}},\
  }\href@noop {} {\bibfield  {journal} {\bibinfo  {journal} {J. Geophys. Res.
  Solid Earth}\ }\textbf {\bibinfo {volume} {101}},\ \bibinfo {pages} {17497}
  (\bibinfo {year} {1996})}\BibitemShut {NoStop}%
\bibitem [{\citenamefont {Teubner}(1990)}]{teubner1990scattering}%
  \BibitemOpen
  \bibfield  {author} {\bibinfo {author} {\bibfnamefont {M.}~\bibnamefont
  {Teubner}},\ }\href@noop {} {\bibfield  {journal} {\bibinfo  {journal} {J.
  Chem. Phys.}\ }\textbf {\bibinfo {volume} {92}},\ \bibinfo {pages} {4501}
  (\bibinfo {year} {1990})}\BibitemShut {NoStop}%
\bibitem [{\citenamefont {Berryman}(1987)}]{berryman1987relationship}%
  \BibitemOpen
  \bibfield  {author} {\bibinfo {author} {\bibfnamefont {J.~G.}\ \bibnamefont
  {Berryman}},\ }\href@noop {} {\bibfield  {journal} {\bibinfo  {journal} {J.
  Math. Phys}\ }\textbf {\bibinfo {volume} {28}},\ \bibinfo {pages} {244}
  (\bibinfo {year} {1987})}\BibitemShut {NoStop}%
\bibitem [{\citenamefont {Doi}(1976)}]{doi1976new}%
  \BibitemOpen
  \bibfield  {author} {\bibinfo {author} {\bibfnamefont {M.}~\bibnamefont
  {Doi}},\ }\href@noop {} {\bibfield  {journal} {\bibinfo  {journal} {J. Phys.
  Soc. Jpn}\ }\textbf {\bibinfo {volume} {40}},\ \bibinfo {pages} {567}
  (\bibinfo {year} {1976})}\BibitemShut {NoStop}%
\bibitem [{\citenamefont {Ma}\ and\ \citenamefont
  {Torquato}(2018)}]{ma2018precise}%
  \BibitemOpen
  \bibfield  {author} {\bibinfo {author} {\bibfnamefont {Z.}~\bibnamefont
  {Ma}}\ and\ \bibinfo {author} {\bibfnamefont {S.}~\bibnamefont {Torquato}},\
  }\href@noop {} {\bibfield  {journal} {\bibinfo  {journal} {Phys. Rev. E}\
  }\textbf {\bibinfo {volume} {98}},\ \bibinfo {pages} {013307} (\bibinfo
  {year} {2018})}\BibitemShut {NoStop}%
\bibitem [{\citenamefont {Prager}(1961)}]{prager1961viscous}%
  \BibitemOpen
  \bibfield  {author} {\bibinfo {author} {\bibfnamefont {S.}~\bibnamefont
  {Prager}},\ }\href@noop {} {\bibfield  {journal} {\bibinfo  {journal} {Phys.
  Fluids}\ }\textbf {\bibinfo {volume} {4}},\ \bibinfo {pages} {1477} (\bibinfo
  {year} {1961})}\BibitemShut {NoStop}%
\bibitem [{\citenamefont {Avellaneda}\ and\ \citenamefont
  {Torquato}(1991)}]{avellaneda1991rigorous}%
  \BibitemOpen
  \bibfield  {author} {\bibinfo {author} {\bibfnamefont {M.}~\bibnamefont
  {Avellaneda}}\ and\ \bibinfo {author} {\bibfnamefont {S.}~\bibnamefont
  {Torquato}},\ }\href@noop {} {\bibfield  {journal} {\bibinfo  {journal}
  {Phys. Fluids A: Fluid Dynamics}\ }\textbf {\bibinfo {volume} {3}},\ \bibinfo
  {pages} {2529} (\bibinfo {year} {1991})}\BibitemShut {NoStop}%
\bibitem [{\citenamefont {Lu}\ and\ \citenamefont
  {Torquato}(1992{\natexlab{a}})}]{lu1992lineal}%
  \BibitemOpen
  \bibfield  {author} {\bibinfo {author} {\bibfnamefont {B.}~\bibnamefont
  {Lu}}\ and\ \bibinfo {author} {\bibfnamefont {S.}~\bibnamefont {Torquato}},\
  }\href@noop {} {\bibfield  {journal} {\bibinfo  {journal} {Phys. Rev. A}\
  }\textbf {\bibinfo {volume} {45}},\ \bibinfo {pages} {922} (\bibinfo {year}
  {1992}{\natexlab{a}})}\BibitemShut {NoStop}%
\bibitem [{\citenamefont {Matheron}(1975)}]{matheron1975random}%
  \BibitemOpen
  \bibfield  {author} {\bibinfo {author} {\bibfnamefont {G.}~\bibnamefont
  {Matheron}},\ }\href@noop {} {\emph {\bibinfo {title} {Random sets and
  integral geometry}}}\ (\bibinfo  {publisher} {Wiley},\ \bibinfo {year}
  {1975})\BibitemShut {NoStop}%
\bibitem [{\citenamefont {Torquato}\ and\ \citenamefont
  {Lu}(1993)}]{torquato1993chord}%
  \BibitemOpen
  \bibfield  {author} {\bibinfo {author} {\bibfnamefont {S.}~\bibnamefont
  {Torquato}}\ and\ \bibinfo {author} {\bibfnamefont {B.}~\bibnamefont {Lu}},\
  }\href@noop {} {\bibfield  {journal} {\bibinfo  {journal} {Phys. Rev. E}\
  }\textbf {\bibinfo {volume} {47}},\ \bibinfo {pages} {2950} (\bibinfo {year}
  {1993})}\BibitemShut {NoStop}%
\bibitem [{\citenamefont {Ho}\ and\ \citenamefont
  {Strieder}(1979)}]{ho1979asymptotic}%
  \BibitemOpen
  \bibfield  {author} {\bibinfo {author} {\bibfnamefont {F.~G.}\ \bibnamefont
  {Ho}}\ and\ \bibinfo {author} {\bibfnamefont {W.}~\bibnamefont {Strieder}},\
  }\href@noop {} {\bibfield  {journal} {\bibinfo  {journal} {J. Chem. Phys}\
  }\textbf {\bibinfo {volume} {70}},\ \bibinfo {pages} {5635} (\bibinfo {year}
  {1979})}\BibitemShut {NoStop}%
\bibitem [{\citenamefont {Tokunaga}(1985)}]{tokunaga1985porous}%
  \BibitemOpen
  \bibfield  {author} {\bibinfo {author} {\bibfnamefont {T.~K.}\ \bibnamefont
  {Tokunaga}},\ }\href@noop {} {\bibfield  {journal} {\bibinfo  {journal} {J.
  Chem. Phys}\ }\textbf {\bibinfo {volume} {82}},\ \bibinfo {pages} {5298}
  (\bibinfo {year} {1985})}\BibitemShut {NoStop}%
\bibitem [{\citenamefont {Thompson}\ \emph {et~al.}(1987)\citenamefont
  {Thompson}, \citenamefont {Katz},\ and\ \citenamefont
  {Krohn}}]{thompson1987microgeometry}%
  \BibitemOpen
  \bibfield  {author} {\bibinfo {author} {\bibfnamefont {A.~H.}\ \bibnamefont
  {Thompson}}, \bibinfo {author} {\bibfnamefont {A.~J.}\ \bibnamefont {Katz}},
  \ and\ \bibinfo {author} {\bibfnamefont {C.~E.}\ \bibnamefont {Krohn}},\
  }\href@noop {} {\bibfield  {journal} {\bibinfo  {journal} {Adv. Phys}\
  }\textbf {\bibinfo {volume} {36}},\ \bibinfo {pages} {625} (\bibinfo {year}
  {1987})}\BibitemShut {NoStop}%
\bibitem [{\citenamefont {Underwood}(1970)}]{underwood1970quantitative}%
  \BibitemOpen
  \bibfield  {author} {\bibinfo {author} {\bibfnamefont {E.~E.}\ \bibnamefont
  {Underwood}},\ }\href@noop {} {\emph {\bibinfo {title} {Quantitative
  stereology}}}\ (\bibinfo {year} {1970})\BibitemShut {NoStop}%
\bibitem [{\citenamefont {Torquato}(1999)}]{torquato1999exact}%
  \BibitemOpen
  \bibfield  {author} {\bibinfo {author} {\bibfnamefont {S.}~\bibnamefont
  {Torquato}},\ }\href@noop {} {\bibfield  {journal} {\bibinfo  {journal} {J.
  Chem. Phys}\ }\textbf {\bibinfo {volume} {111}},\ \bibinfo {pages} {8832}
  (\bibinfo {year} {1999})}\BibitemShut {NoStop}%
\bibitem [{\citenamefont {Jiao}\ \emph {et~al.}(2007)\citenamefont {Jiao},
  \citenamefont {Stillinger},\ and\ \citenamefont
  {Torquato}}]{jiao2007modeling}%
  \BibitemOpen
  \bibfield  {author} {\bibinfo {author} {\bibfnamefont {Y.}~\bibnamefont
  {Jiao}}, \bibinfo {author} {\bibfnamefont {F.~H.}\ \bibnamefont
  {Stillinger}}, \ and\ \bibinfo {author} {\bibfnamefont {S.}~\bibnamefont
  {Torquato}},\ }\href@noop {} {\bibfield  {journal} {\bibinfo  {journal}
  {Phys. Rev. E}\ }\textbf {\bibinfo {volume} {76}},\ \bibinfo {pages} {031110}
  (\bibinfo {year} {2007})}\BibitemShut {NoStop}%
\bibitem [{\citenamefont {Jiao}\ \emph {et~al.}(2009)\citenamefont {Jiao},
  \citenamefont {Stillinger},\ and\ \citenamefont
  {Torquato}}]{jiao2009superior}%
  \BibitemOpen
  \bibfield  {author} {\bibinfo {author} {\bibfnamefont {Y.}~\bibnamefont
  {Jiao}}, \bibinfo {author} {\bibfnamefont {F.~H.}\ \bibnamefont
  {Stillinger}}, \ and\ \bibinfo {author} {\bibfnamefont {S.}~\bibnamefont
  {Torquato}},\ }\href@noop {} {\bibfield  {journal} {\bibinfo  {journal}
  {Proc. Natl. Acad. Sci. U.S.A.}\ }\textbf {\bibinfo {volume} {106}},\
  \bibinfo {pages} {17634} (\bibinfo {year} {2009})}\BibitemShut {NoStop}%
\bibitem [{\citenamefont {Chen}\ \emph {et~al.}(2015)\citenamefont {Chen},
  \citenamefont {Li},\ and\ \citenamefont {Jiao}}]{chen2015dynamic}%
  \BibitemOpen
  \bibfield  {author} {\bibinfo {author} {\bibfnamefont {S.}~\bibnamefont
  {Chen}}, \bibinfo {author} {\bibfnamefont {H.}~\bibnamefont {Li}}, \ and\
  \bibinfo {author} {\bibfnamefont {Y.}~\bibnamefont {Jiao}},\ }\href@noop {}
  {\bibfield  {journal} {\bibinfo  {journal} {Phys. Rev. E}\ }\textbf {\bibinfo
  {volume} {92}},\ \bibinfo {pages} {023301} (\bibinfo {year}
  {2015})}\BibitemShut {NoStop}%
\bibitem [{\citenamefont {Karsanina}\ and\ \citenamefont
  {Gerke}(2018)}]{karsanina2018hierarchical}%
  \BibitemOpen
  \bibfield  {author} {\bibinfo {author} {\bibfnamefont {M.~V.}\ \bibnamefont
  {Karsanina}}\ and\ \bibinfo {author} {\bibfnamefont {K.~M.}\ \bibnamefont
  {Gerke}},\ }\href@noop {} {\bibfield  {journal} {\bibinfo  {journal} {Phys.
  Rev. Lett}\ }\textbf {\bibinfo {volume} {121}},\ \bibinfo {pages} {265501}
  (\bibinfo {year} {2018})}\BibitemShut {NoStop}%
\bibitem [{\citenamefont {{\v{C}}apek}(2018)}]{vcapek2018importance}%
  \BibitemOpen
  \bibfield  {author} {\bibinfo {author} {\bibfnamefont {P.}~\bibnamefont
  {{\v{C}}apek}},\ }\href@noop {} {\bibfield  {journal} {\bibinfo  {journal}
  {Transport Porous Med}\ }\textbf {\bibinfo {volume} {125}},\ \bibinfo {pages}
  {59} (\bibinfo {year} {2018})}\BibitemShut {NoStop}%
\bibitem [{\citenamefont {Li}\ \emph {et~al.}(2018)\citenamefont {Li},
  \citenamefont {Zhang}, \citenamefont {Zhao}, \citenamefont {Burkhart},
  \citenamefont {Brinson},\ and\ \citenamefont {Chen}}]{li2018transfer}%
  \BibitemOpen
  \bibfield  {author} {\bibinfo {author} {\bibfnamefont {X.}~\bibnamefont
  {Li}}, \bibinfo {author} {\bibfnamefont {Y.}~\bibnamefont {Zhang}}, \bibinfo
  {author} {\bibfnamefont {H.}~\bibnamefont {Zhao}}, \bibinfo {author}
  {\bibfnamefont {C.}~\bibnamefont {Burkhart}}, \bibinfo {author}
  {\bibfnamefont {L.~C.}\ \bibnamefont {Brinson}}, \ and\ \bibinfo {author}
  {\bibfnamefont {W.}~\bibnamefont {Chen}},\ }\href@noop {} {\bibfield
  {journal} {\bibinfo  {journal} {Sci. Rep.}\ }\textbf {\bibinfo {volume}
  {8}},\ \bibinfo {pages} {1} (\bibinfo {year} {2018})}\BibitemShut {NoStop}%
\bibitem [{\citenamefont {Pant}\ \emph {et~al.}(2015)\citenamefont {Pant},
  \citenamefont {Mitra},\ and\ \citenamefont {Secanell}}]{pant2015multigrid}%
  \BibitemOpen
  \bibfield  {author} {\bibinfo {author} {\bibfnamefont {L.~M.}\ \bibnamefont
  {Pant}}, \bibinfo {author} {\bibfnamefont {S.~K.}\ \bibnamefont {Mitra}}, \
  and\ \bibinfo {author} {\bibfnamefont {M.}~\bibnamefont {Secanell}},\
  }\href@noop {} {\bibfield  {journal} {\bibinfo  {journal} {Phys. Rev. E}\
  }\textbf {\bibinfo {volume} {92}},\ \bibinfo {pages} {063303} (\bibinfo
  {year} {2015})}\BibitemShut {NoStop}%
\bibitem [{\citenamefont {Gerke}\ \emph {et~al.}(2019)\citenamefont {Gerke},
  \citenamefont {Karsanina},\ and\ \citenamefont
  {Katsman}}]{gerke2019calculation}%
  \BibitemOpen
  \bibfield  {author} {\bibinfo {author} {\bibfnamefont {K.~M.}\ \bibnamefont
  {Gerke}}, \bibinfo {author} {\bibfnamefont {M.~V.}\ \bibnamefont
  {Karsanina}}, \ and\ \bibinfo {author} {\bibfnamefont {R.}~\bibnamefont
  {Katsman}},\ }\href@noop {} {\bibfield  {journal} {\bibinfo  {journal} {Phys.
  Rev. E}\ }\textbf {\bibinfo {volume} {100}},\ \bibinfo {pages} {053312}
  (\bibinfo {year} {2019})}\BibitemShut {NoStop}%
\bibitem [{\citenamefont {Torquato}(2020)}]{torquato2020predicting}%
  \BibitemOpen
  \bibfield  {author} {\bibinfo {author} {\bibfnamefont {S.}~\bibnamefont
  {Torquato}},\ }\href@noop {} {\bibfield  {journal} {\bibinfo  {journal} {Adv.
  Water Resour.}\ }\textbf {\bibinfo {volume} {140}},\ \bibinfo {pages}
  {103565} (\bibinfo {year} {2020})}\BibitemShut {NoStop}%
\bibitem [{\citenamefont {Jiao}\ \emph {et~al.}(2008)\citenamefont {Jiao},
  \citenamefont {Stillinger},\ and\ \citenamefont
  {Torquato}}]{jiao2008modeling}%
  \BibitemOpen
  \bibfield  {author} {\bibinfo {author} {\bibfnamefont {Y.}~\bibnamefont
  {Jiao}}, \bibinfo {author} {\bibfnamefont {F.~H.}\ \bibnamefont
  {Stillinger}}, \ and\ \bibinfo {author} {\bibfnamefont {S.}~\bibnamefont
  {Torquato}},\ }\href@noop {} {\bibfield  {journal} {\bibinfo  {journal}
  {Phys. Rev. E}\ }\textbf {\bibinfo {volume} {77}},\ \bibinfo {pages} {031135}
  (\bibinfo {year} {2008})}\BibitemShut {NoStop}%
\bibitem [{\citenamefont {Lu}\ and\ \citenamefont
  {Torquato}(1992{\natexlab{b}})}]{lu1992lineal2}%
  \BibitemOpen
  \bibfield  {author} {\bibinfo {author} {\bibfnamefont {B.}~\bibnamefont
  {Lu}}\ and\ \bibinfo {author} {\bibfnamefont {S.}~\bibnamefont {Torquato}},\
  }\href@noop {} {\bibfield  {journal} {\bibinfo  {journal} {Phys. Rev. A}\
  }\textbf {\bibinfo {volume} {45}},\ \bibinfo {pages} {7292} (\bibinfo {year}
  {1992}{\natexlab{b}})}\BibitemShut {NoStop}%
\bibitem [{\citenamefont {Hansen}\ and\ \citenamefont
  {McDonald}(1986)}]{hansen1990theory}%
  \BibitemOpen
  \bibfield  {author} {\bibinfo {author} {\bibfnamefont {J.-P.}\ \bibnamefont
  {Hansen}}\ and\ \bibinfo {author} {\bibfnamefont {I.~R.}\ \bibnamefont
  {McDonald}},\ }\href@noop {} {\emph {\bibinfo {title} {Theory of Simple
  Liquids}}}\ (\bibinfo  {publisher} {Academic Press},\ \bibinfo {year}
  {1986})\BibitemShut {NoStop}%
\bibitem [{\citenamefont {Torquato}\ and\ \citenamefont
  {Stell}(1985)}]{torquato1985microstructure}%
  \BibitemOpen
  \bibfield  {author} {\bibinfo {author} {\bibfnamefont {S.}~\bibnamefont
  {Torquato}}\ and\ \bibinfo {author} {\bibfnamefont {G.}~\bibnamefont
  {Stell}},\ }\href@noop {} {\bibfield  {journal} {\bibinfo  {journal} {J.
  Chem. Phys}\ }\textbf {\bibinfo {volume} {82}},\ \bibinfo {pages} {980}
  (\bibinfo {year} {1985})}\BibitemShut {NoStop}%
\bibitem [{\citenamefont {Torquato}(1995)}]{torquato1995nearest}%
  \BibitemOpen
  \bibfield  {author} {\bibinfo {author} {\bibfnamefont {S.}~\bibnamefont
  {Torquato}},\ }\href@noop {} {\bibfield  {journal} {\bibinfo  {journal}
  {Phys. Rev. E}\ }\textbf {\bibinfo {volume} {51}},\ \bibinfo {pages} {3170}
  (\bibinfo {year} {1995})}\BibitemShut {NoStop}%
\bibitem [{\citenamefont {Gommes}\ \emph {et~al.}(2012)\citenamefont {Gommes},
  \citenamefont {Jiao},\ and\ \citenamefont {Torquato}}]{gommes2012density}%
  \BibitemOpen
  \bibfield  {author} {\bibinfo {author} {\bibfnamefont {C.~J.}\ \bibnamefont
  {Gommes}}, \bibinfo {author} {\bibfnamefont {Y.}~\bibnamefont {Jiao}}, \ and\
  \bibinfo {author} {\bibfnamefont {S.}~\bibnamefont {Torquato}},\ }\href@noop
  {} {\bibfield  {journal} {\bibinfo  {journal} {Phys. Rev. Lett}\ }\textbf
  {\bibinfo {volume} {108}},\ \bibinfo {pages} {080601} (\bibinfo {year}
  {2012})}\BibitemShut {NoStop}%
\end{thebibliography}
\end{document}